 \documentclass[prd,aps,showpacs,preprintnumbers,amsmath,amssymb,nofootinbib,eqsecnum, preprintnumbers]{revtex4}

\usepackage{graphicx}
\usepackage{epsf}
\usepackage{amsmath}
\usepackage{epstopdf}
\usepackage{bm}
\usepackage{color}
\usepackage{tabularx}
\usepackage{enumitem}
\usepackage{float}
\usepackage{array,booktabs}
\usepackage{footnote}
\usepackage{threeparttable}
\usepackage{graphicx}
\usepackage{hyperref}
\usepackage{amssymb,epsf}
\usepackage{latexsym}
\usepackage{epstopdf}
\usepackage{epsfig}

\usepackage[normalem]{ulem}
\useunder{\uline}{\ul}{}

\begin{document}

\title{Nonlinearly Charged Black Hole Chemistry with Massive Gravitons \\ in the Grand Canonical Ensemble}

\author{Ali Dehghani$^{1,2}$\footnote{email address: ali.dehghani.phys@gmail.com} and
Seyed Hossein Hendi$^{1,2,3}$\footnote{email address:
hendi@shirazu.ac.ir (Corresponding author)}}

\affiliation{
 $^1$Department of Physics, School of Science, Shiraz University, Shiraz 71454, Iran \\
 $^2$Biruni Observatory, School of Science, Shiraz University, Shiraz 71454, Iran    \\
 $^{3}$Canadian Quantum Research Center 204-3002 32 Ave Vernon, BC V1T 2L7 Canada    }

\date{\today}

\begin{abstract}
In the context of Black Hole Chemistry (BHC), holographic phase
transitions of asymptotically anti-de Sitter (AdS) charged
topological black holes (TBHs) in massive gravity coupled to Power
Maxwell Invariant (PMI) electrodynamics are discussed in the grand
canonical (fixed $U(1)$ potential, $\Phi$) ensemble. Considering
the higher-order of graviton's self-interactions of (dRGT) massive
gravitational field theory in arbitrary dimensions, we derive
exact TBH solutions. We also calculate the explicit form of the
on-shell action and the associated thermodynamic quantities in the
Grand Canonical Ensemble (GCE). Besides, we examine the validity
of the first law of thermodynamics and Smarr relation in the
extended phase space thermodynamics. Regarding this model, we show
that a) a van der Waals (vdW) behavior takes place in $d \ge 4$
dimensions, b) a typical reentrant phase transition can happen in
$d \ge 5$, and c) both anomalous vdW and triple point phenomena
are recognized in $d \ge 6$. Accordingly, we find that the
obtained results are quite different from their counterparts in
the GCE of Maxwell-massive gravity. Furthermore, we briefly study
the critical behaviors of higher-dimensional TBHs with a
conformally invariant Maxwell source as a specific subclass of our
solutions for Einstein's theory and the massive gravity in various
ensembles.
\end{abstract}

 \maketitle

\section{Introduction} \label{sec:intro}

The subject of Black Hole Chemistry (BHC) \cite{CQG2017Review,
Karch2015} has received much attention recently due to the
similarity between specific holographic phase transitions and
those counterparts in the real world. In this regard, van der
Waals (vdW) behavior \cite{KubiznakMann2012}, reentrant phase
transition (RPT) with the associated large/intermediate/large
black hole (LBH/IBH/LBH) phase transition
\cite{Mann2012Gunasekaran,Mann2012Altamirano}, triple point
phenomenon with the associated small/intermediate/large black hole
(SBH/IBH/LBH) phase transition \cite{Mann2014TriplePoint}, and
$\lambda$-line black hole phase transition
\cite{HennigarMann2017PRL} have been observed in various
asymptotically AdS black hole configurations. Developments of BHC
are deeply rooted in extending the usual thermodynamic phase space
(known as the extended phase space
\cite{Kastor2009CQG,Kastor2010LovelockSmarr,Kastor2018}), but
criticality may take place in non-extended phase spaces as well
(for these alternatives, see
\cite{EmparanChamblin1999a,EmparanChamblin1999b,Caldarelli2000KerrNewmanAdS,Fernando(2006)BI-AdS,Mo2015NonExtendedLovelock,Cai2015,Dehyadegari2020}
and references therein). However, in the extended phase space,
there is an exact identification between intensive/extensive
quantities of AdS black holes and fluid systems
\cite{KubiznakMann2012} which is absent in non-extended phase
space. Thus, it seems to be an important issue to consider the
extended phase space thermodynamics and one can learn many things
from it along different lines
\cite{CQG2017Review,Dolan2011CQG1,Dolan2011CQG2,Dolan2011PRD}. At
present, a great deal of research has devoted to exploring the
$P-v$ criticality and phase structures of different charged-AdS
black hole configurations
\cite{Wei2012,Cai2013GaussBonnet,HendiVahidinia2013,Frassino2014,MannGalaxies2014,DolanMann2014Lovelock,BI-AdSBH-PV2014,PV2014Lovelock,PV2015LovelockBI-Belhaj,PVmassive2015PRD,Hendi2017Mann-PRD,RPT-dRGTmassive-2017,PV-BlackBranes-2017Hennigar,Mir2019a,Mir2019b,EPJC2019,PRD2020,CQG2020,PV-2020-Triple-MassiveGravity,PV-2019-Einstein-Horndeski,PV-2020-BraneworldBHs,Gue2020},
exhibiting a vast variety of critical phenomena sometimes close to
real-world experiments.

Elementally, $P-v$ criticality can take place in both the
canonical and grand canonical ensembles. However, in black hole
physics, the case of GCE (fixed $U(1)$ potential, $\Phi$) does not
generally consider as an \textit{interesting} case since only
canonical ensemble phase transitions (the ensemble with fixed
$U(1)$ charge, $Q$, at infinity) are observed for charged-AdS
black holes in Einstein gravity \cite{KubiznakMann2012}. The same
statement holds for rotating-AdS black holes; black hole phase
transition takes place in the canonical ensemble with fixed
angular momentum ($J$) at infinity but not in the GCE with fixed
angular velocity ($\Omega$) at infinity \cite{MannGalaxies2014}.
Nevertheless, this statement is no longer true in modified
theories of gravity and, in some cases such as charged TBHs of
massive gravity, the grand canonical phase structure is even
richer than the canonical ensemble \cite{CQG2020}. So, modified
gravity theories provide a more complex environment for TBHs which
leads to richer phase structures and subsequently more
possibilities for the investigation of BHC. On the other hand,
modified gravities provide a good framework to examine BHC in
various ensembles.

Among various candidates of the modified theories of gravitation,
massive gravity theories (see some excellent reviews in Refs.
\cite{Hinterbichler2012Review,deRhamREVIEW2014}) have received
increased attention lately due to a wide variety of motivations
such as explaining the current observations related to dark matter
\cite{Schmidt-May2016a-DM,Schmidt-May2016b-DM}, describing the
accelerating expansion of the universe without requiring any dark
energy component \cite{Akrami2013,Akrami2015}, etc. The dRGT
version of massive gravity theory
\cite{deRham2010Gabadadze,dRGT2011} has consistently resolved the
long-standing problem of gravitational fields with spin-2 massive
mediators (i.e., gravitons with a tiny mass, $m_g$), avoiding
Boulware-Deser ghost \cite{BDghost1972,HassanRosen2012PRL} and
vDVZ discontinuity \cite{vDVZ1970a,vDVZ1970b} (which appears in
Fierz-Pauli theory \cite{FierzPauli1939}). This theory requires a
fiducial reference metric ($f_{\mu \nu}$) in addition to the
dynamical metric ($g_{\mu \nu}$) in order to define a mass term
for graviton by introducing some non-derivative potential terms
(${\cal U}_{i}$). Elementally, each choice of reference metric
leads to a special family of massive gravity \cite{MG2012JHEP} and
in several studies, it is proved that the dRGT massive gravity is
ghost-free by assuming various reference metrics such as Minkowski
and degenerate (singular) reference metrics
\cite{HassanRosen2012PRL,Alberte2013PRD,Alberte2015PRD,ReducedMG-2016-GhostFree,TQDo2016massive,TQDo2016bigravity}.
Therefore, considering different possibilities for the reference
metric could lead to a variety of new solutions. Cosmological
studies of massive gravity have primarily focused on the Minkowski
reference metric
\cite{MassiveCosmologies2011,Chamseddine2011Volkov,TQDo2016massive,TQDo2016bigravity,deRham2017}.
On the other hand, there have been found a number of
asymptotically flat and (A)dS black hole solutions in the context
of massive gravity which relied either on the flat (Minkowski)
reference metric or on a degenerate (spatial) and singular
reference metric
\cite{Koyama2011(PRL),Nieuwenhuizen2011(PRD),Gruzinov2011Mirbabayi,Berezhiani2012,Rosen2017BHs,Vegh2013,Cai2015}.
A particularly interesting case is the AdS black hole solutions
with the degenerate (spatial) reference metric which is singular
and has important applications in gauge/gravity duality
\cite{Vegh2013} as well as BHC \cite{PRD2020}. This class of
solutions are dual to homogenous and isotropic condensed matter
systems \cite{Vegh2013,BlakeTong2013,Davison2013} and can describe
different phases of matter with broken translational symmetry
\cite{Alberte2015PRD,Baggioli2015PRL,Baggioli2016JHEP,Baggioli2018PRL}.
Within the BHC framework, a new range of phase transitions can be
found for this class of AdS black holes \cite{PRD2020}.

Concentrating on the subject of BHC, it is found that
higher-dimensional TBHs in dRGT massive gravity can mimic the
critical behavior of everyday substances in nature without the
inclusion of any extra or unusual matter fields in the bulk action
\cite{PRD2020}. In fact, a finite number of graviton's
self-interaction potentials (${\cal U}_{i}$) exist in each
dimension and, naturally, more potential terms contribute to
higher dimensions. So, in principle, there are additional degrees
of freedom in higher dimensions that enrich the thermodynamic
phase space of TBHs. In Ref. \cite{CQG2020}, a generalization of
this study for charged TBHs has been performed in both the
canonical (fixed charge, $Q$) and grand canonical (fixed
potential, $\Phi$) ensembles showing that the nature of phase
transitions depends on the spacetime dimensions and also the
ensemble one is dealing with. As indicated in Ref. \cite{CQG2020},
for charged TBHs within the framework of Maxwell-massive gravity,
more interesting (critical) phenomena are observed in the GCE in
comparison with the canonical ensemble. In the present paper, we
intend to study the chemistry of TBHs in a more complex
environment via massive gravity coupled with a nonlinear $U(1)$
sector arisen from the power Maxwell invariant (PMI)
electrodynamics \cite{HassaineMartinez2007,HassaineMartinez2008}
in the GCE.\footnote{In Ref. \cite{Canonical-PMI-massive}, the
phase space structure of this model in the canonical ensemble has
been investigated in detail. In Sect. \ref{sec:comparison} of the
present paper, we will compare the outcomes of criticality for
both ensembles.} As will be evident in the present work, a massive
spin-2 field Lagrangian minimally coupled to a PMI $U(1)$ gauge
field causes a range of novel black hole phase transitions in both
the canonical and grand canonical ensembles. In addition, this
model up to four graviton's self-interactions (${\cal U}_4$) has
already been explored in Ref. \cite{Dehyadegari2017MGPMI} in order
to investigate the electrical transport behavior of the dual field
theory in the presence of a PMI gauge field for the 4- and
5-dimensional black brane solutions.

The theory of PMI electrodynamics as a matter source for AdS black
holes has resulted in some interesting consequences
\cite{HassaineMartinez2007,HassaineMartinez2008,Hendi2009Sedehi,Hendi2010PMI,HendiVahidinia2013,Dehyadegari2017MGPMI,Dehyadegari2020}.
The Lagrangian density of PMI electrodynamics is given by a power
of the Maxwell Lagrangian as
\begin{equation} \label{PMI Lagrangian}
{{\cal L}_{{\rm{PMI}}}} = {( - {\cal F})^s},
\end{equation}
in which ${\cal F} = {F_{\mu \nu }}{F^{\mu \nu }}$, $s$ is the
nonlinearity parameter and the Faraday tensor ($F_{\mu \nu}$) is
defined as ${F_{\mu \nu }} = 2 {\partial _{[\mu }}{A_{\nu ]}}$. In
the limit $s \to 1$, the standard results of $U(1)$ Maxwell
electrodynamics can be recovered simply. There exist several
motivations for considering such a model of nonlinear
electrodynamics, listed below:

\begin{itemize}
    \item This theory allows us to
    examine various deviations from the Coulomb law
    \cite{HassaineMartinez2007,HassaineMartinez2008}. These deviations can be realized by comparing the electric potential in Maxwell's electrodynamics and PMI theory, i.e.,
    \begin{equation} \label{PMI potential law}
    {\rm{Maxwell}}: V \propto \frac{1}{{{r^{d - 3}}}} \longrightarrow {\rm{PMI}}: V \propto\frac{1}{{{r^{\frac{{d - 2s - 1}}{{2s - 1}}}}}}.
    \end{equation}
    The range of allowed deviations from the Coulomb law is bounded by demanding a
    finite value for the electric potential at infinity and also by implementing the causality and
    unitarity principles to the PMI Lagrangian (see more details in Sec. \ref{sec:II}), yielding $1 \leq s < \frac{d-1}{2}$. Using the PMI field equations in the vacuum, it is straightforward to show that the relative permittivity and the relative permeability are given by
    \begin{equation} \label{relative permeability}
    \frac{\varepsilon }{{{\varepsilon _0}}} = \frac{{{\mu _0}}}{\mu } = s{( - {\cal F})^{s - 1}}= s{({E^2} - {c^2}{B^2})^{s - 1}},
    \end{equation}
    which clearly indicates the polarization of vacuum (this result is different from that of Born-Infeld (BI) theory since their effective Lagrangians have different functional forms.) In the limit $s \to 1$, Eq. (\ref{relative permeability}) reduces to the case of the absence of vacuum polarization effects (${\varepsilon }/{{{\varepsilon _0}}} = {{{\mu _0}}}/{\mu } =1$), which is a familiar result in the Maxwell theory. In comparison with effective field theories inspired by QED \cite{Heisenberg1936,NED-QED-a,NED-QED-b,NED-QED-c},
    this vacuum polarization can be justified by the presence of virtual pair particles as effective dipoles in vacuum
    (the well-known screening effect). In 4-dimensions, for infinitesimal deviations from Coulomb law, the PMI potential law \ref{PMI potential law} takes the form $V \propto \frac{1}{r^{1-\epsilon}}$ where $\epsilon $ is always positive due to the causality and
    unitarity conditions. As an example, for $r \gg 1$, we find that the PMI potential is slightly larger than the Coulomb law, qualitatively (but not quantitatively) in agreement with the QED Uehling potential \cite{Peskin}. Besides, we should note that PMI is a classical model and setting an energy scale for classical theories is ambiguous. However, with a
    specific series expansion of the PMI Lagrangian (\ref{PMI Lagrangian}), one can obtain a BI-type Lagrangian (see the next item) which simulates the results of QED
    the same as the Euler-Heisenberg theory.
\end{itemize}
\begin{itemize}
    \item The series expansion of the PMI Lagrangian at the unknown constant (${\cal F}_0$) yields \cite{HendiVahidinia2013}
    \begin{eqnarray} \label{PMI expansion}
    {{\cal L}_{{\rm{PMI}}}}({\cal F}) &=& \sum\limits_{n = 0}^\infty  {\frac{{{\cal L}_{{\rm{PMI}}}^{\,(n)}({{\cal F}_0})}}{{n!}}} {\left( {{\cal F} - {{\cal F}_0}} \right)^n} \\ \nonumber
    &\simeq& {a_1}{\cal F} + (s - 1) \big[ {{a_0} + {a_2}{{( - {\cal F})}^2} + {a_3}{{( - {\cal F})}^3} + ...} \big],
    \end{eqnarray}
    where we have set ${a_1}=-1$ to recover the standard linear Lagrangian ($- \cal F$) in the Maxwell limit $s \to 1$.
    The other constants $a_i$'s depend on $s$ and ${\cal F}_0$. So, up to the second order approximation, we end up with a
    quadratic Maxwell invariant (${\cal F}^2$) in addition to the Maxwell Lagrangian in a way reminiscent of BI-type theories of nonlinear electrodynamics since BI-type theories generally have a series expansion as
    ${{\cal L}_{\rm{BI}}}\simeq -{\cal F}+{b_1}{\cal F}^2+O({\cal F}^4)$ \cite{Hendi2012}.
    Perhaps, the most important motivation for considering the BI-type nonlinear theories of electrodynamics comes from the fact that one loop QED corrections can be effectively understood via
    the quadratic Maxwell invariant term of these theories \cite{EFT-NED-QED,Shabad2015}.
    As seen, this is also valid for the expansion of the PMI theory (\ref{PMI expansion}) up to the second order approximation. In fact, there exist a direct analogy between the relative permittivity and the relative permeability (in BI-type nonlinear models) and the vacuum permittivity and permeability tensors in QED. In such cases, one can establish a connection between the (classical electrodynamics) nonlinear parameter and the QED coefficient in the vacuum permittivity and permeability tensors  \cite{Jackson}. For these reasons, nonlinear models of electrodynamics can classically simulate some quantum mechanical effects of QED. In addition, it is shown that nonlinear electrodynamics with certain assumptions contains low-energy QED as a special case  \cite{Fouch}. In conclusion, at scales where loop QED corrections have significant contribution,
    such modifications can be classically considered without
    requiring any quantum mechanical treatment.
\end{itemize}
\begin{itemize}
        \item The Maxwell action enjoys conformal
        invariance in 4-dimensions but it does not
        possess this symmetry in higher dimensions. However, the PMI action extends the
        conformal invariance in higher dimensions if the power is chosen
        as $s=d/4$ \cite{HassaineMartinez2007}. Therefore, with this choice, the corresponding energy-momentum tensor will be traceless in arbitrary dimensions.
        In this way, it is possible to have a conformally invariant Maxwell source for black hole systems in arbitrarily higher dimensions.
\end{itemize}

\begin{itemize}
    \item It has been recently indicated that, under certain conditions, the PMI theory with the Lagrangian density
    (\ref{PMI Lagrangian}) can also remove the singularity of the electric field of point-like charges \cite{Eslampanah2021}.
\end{itemize}
\begin{itemize}
    \item   The low energy limit of ${E_8} \times {E_8} $ heterotic string theory reduces to an effective field theory including a quadratic
    Maxwell invariant term as well as the Maxwell Lagrangian \cite{Gross1987}. Moreover, it has been demonstrated that electromagnetic fields on the
    world-volumes of D-branes are governed by BI-type theories \cite{ZwiebachStringBook}. In conclusion, inspired by string theory models,
    one can replace the conventional Maxwell Lagrangian with the Lagrangian of PMI theory which essentially involves higher-derivative
    corrections of the $U(1)$ gauge field. For this reason, such unconventional coupling to nonlinear source of electrodynamics has been
    already investigated in the context of AdS/CFT holographic superconductors \cite{AdS/CFT-PMI-a,AdS/CFT-PMI-b,AdS/CFT-PMI-c,AdS/CFT-PMI-d}.
\end{itemize}
Considering the massive gravity on AdS space in the presence of
PMI electromagnetic field leads to a new class of TBHs exhibiting
a range of phase transitions in both the canonical and the grand
canonical ensembles. The nonlinear effect of electromagnetic field
arising from the PMI theory introduces new interesting features
for black hole thermodynamics beyond the Einstein-Maxwell gravity.
Naturally, this property enriches the extended thermodynamic phase
space, as proved for AdS black holes in Einstein gravity coupled
with PMI electrodynamics in which spherically symmetric black
holes exhibit $P-v$ criticality in both the canonical and the
grand canonical ensembles \cite{HendiVahidinia2013} (Note that the
GCE phase transition for charged-AdS black holes is absent in
Einstein-Maxwell gravity \cite{KubiznakMann2012}.) This theory
also allows us to explore the extended phase space thermodynamics
of charged TBHs with a conformally invariant Maxwell (CIM) source,
which has not yet been studied. As a matter of fact, this leads to
AdS black hole spacetimes with the property of the inverse square
electric field (the so-called Coulomb law) in higher dimensions
\cite{HassaineMartinez2007,HassaineMartinez2008} and this is the
case for the massive gravity theory, as will be discussed in this
paper. It should be noted that the outcomes of Einstein-Maxwell
gravity is naturally recovered by taking the limits $m_g \to 0$
(massless graviton limit) and $s \to 1$ (the Maxwell limit).\\

The chemistry of TBHs in modified gravity coupled to a nonlinear
electromagnetic source is a challenging task and requires further
attention since the thermodynamics of TBHs takes a more
complicated structure. Therefore, we study the subjects of BHC and
the extended phase space thermodynamics of charged TBHs in the
context of PMI-massive gravity in detail. To do so, this paper is
organized as follows: in Sec. \ref{sec:II}, we review the theory
of PMI-massive gravity and find the exact charged TBH solutions.
In Sec. \ref{secIII}, we obtain the semi-classical partition
function of TBHs with the fixed potential at infinity as a
boundary condition. Then, in Sec. \ref{secIV}, we extract the
thermodynamic quantities using the grand partition function and
show that the first law of extended thermodynamics is satisfied.
In Sec. \ref{secV} and Sec. \ref{sec:comparison}, we analyze the
holographic phase transitions in detail and compare them with the
results of the canonical ensemble, respectively. Afterward, in
Sec. \ref{sec:Massive-CIM}, we investigate briefly the critical
behaviors of charged TBHs in Einstein and massive gravity theories
coupled with a CIM source. In Sec. \ref{sec:exponents}, we compute
the critical exponents for the obtained TBHs in both ensembles.
Finally, in Sec. \ref{sec:conclusion}, we summarize the main
results.

\section{PMI-massive gravity} \label{sec:II}

Since in the presence of negative cosmological constant,
gravitational field equations admit black hole solutions of
constant (positive, zero and negative) curvature for the event
horizon, we consider such TBH spacetimes with various horizon's
geometries.  But, Ricci flat and hyperbolically symmetric black
holes do not admit phase transitions in Einstein gravity, so one
needs additional degrees of freedom which enrich the thermodynamic
phase space. To do so, one can define an explicit mass term for
gravitons, and then see that all kinds of TBHs can experience
critical behavior and phase transition exactly in the same way due
to the massive graviton's self-interaction potentials
\cite{PRD2020}. In addition, we intend to consider a more general
(nonlinear) electrodynamics in terms of a power of the Maxwell
invariant \cite{HassaineMartinez2007,HassaineMartinez2008}, more
specifically with the form $(-{\cal F})^s$, which recovers the
outcomes of Maxwell electrodynamics (if $s=1$) and also enjoys
conformal invariance in higher dimensions if the deformation
parameter $s$ is chosen properly, i.e., $s=d/4$. For this purpose,
we select the theory of PMI electrodynamics as the nonlinear
matter source with the Lagrangian (\ref{PMI Lagrangian}). This
nonlinear Lagrangian allows us to examine various deviations from
the Coulomb law.

In this paper, we are dealing with the static charged topological
black holes without a magnetic field. Assuming this, only the
electrostatic field contributes in the Faraday tensor, ($F_{\mu
\nu }$), which leads to ${\cal F} = {F_{\mu \nu }}{F^{\mu \nu }} =
- 2{\left({\frac{dh(r)}{dr}}\right)^2}$ in arbitrary dimensions
($h(r)$ is the scalar potential). So, the Lagrangian (\ref{PMI
Lagrangian}) is always positive definite in its domain. However,
one should take care of the causality and unitarity criteria in
different models of nonlinear electrodynamics. By implementing the
causality and unitarity criteria to the local effective action of
nonlinear electrodynamics, the authors of Ref. \cite{Shabad2011}
have found some requirements in terms of some inequality relations
for the Lagrangian density as
\begin{equation}
\frac{{\partial {\cal L}}}{{\partial {\cal F}}} \le 0\,,\,\,
\frac{{{\partial ^2}{\cal L}}}{{\partial {{\cal F}^2}}} \ge
0\,\,\,{\rm{and}}\,\,\,\frac{{\partial {\cal L}}}{{\partial {\cal
F}}} + 2{\cal F}\frac{{{\partial ^2}{\cal L}}}{{\partial {{\cal
F}^2}}} \le 0\,.
\end{equation}
These conditions are equivalent to the positive convexity of the
effective Lagrangian of nonlinear electrodynamics. For the PMI
model, the above requirements are satisfied as
\begin{equation}
\frac{{\partial {{\cal L}_{{\rm{PMI}}}}}}{{\partial {\cal F}}} =  - s{( - {\cal F})^{s - 1}} \le 0,
\end{equation}
\begin{equation}
\frac{{{\partial ^2}{{\cal L}_{{\rm{PMI}}}}}}{{\partial {{\cal F}^2}}} = s(s - 1){( - {\cal F})^{s - 2}} \ge 0,
\end{equation}
and
\begin{equation}
\frac{{\partial {{\cal L}_{{\rm{PMI}}}}}}{{\partial {\cal F}}} + 2{\cal F}\frac{{{\partial ^2}{{\cal L}_{{\rm{PMI}}}}}}{{\partial {{\cal F}^2}}} =  - s(2s - 1){( - {\cal F})^{s - 1}} \le 0,
\end{equation}
which prove that PMI theory satisfies the causality and unitarity
conditions for $s \ge 1$.

As stated, we are going to look for new phenomena that are not
observed in the Einstein-Maxwell, Einstein-PMI and Maxwell-massive
gravitational systems. Motivated by this and also following the
aforementioned motivations in Sec. \ref{sec:intro}, we construct
the full nonlinear theory of dRGT massive gravity
\cite{deRham2010Gabadadze,dRGT2011} minimally coupled with PMI
electrodynamics and this is exactly what we need for our purpose.
Hence, the bulk action for this theory on AdS is written as
\begin{equation} \label{bulk action}
{{\cal I}_{\rm{b}}} =  - \frac{1}{{16\pi {G_d}}}\int_{\cal M}
{{d^d}x\sqrt { - g} \Big[R - 2\Lambda  + m_g^2\sum\limits_{i =
1}^{d - 2} {{c_i}{{\cal U}_i}(g,f)}  + {{{\cal L}_{{\rm{PMI}}}}}
\Big]},
\end{equation}
where $\Lambda$ is the cosmological constant related to the AdS
radius ($\ell$) by $\Lambda  =- \frac{(d-1)(d-2)}{2{\ell ^2}}$. In
the bulk action (\ref{bulk action}), $m_g$ is the graviton's mass
parameter, $c_i$'s are arbitrary constants and ${\cal U}_i$'s are
graviton's self-interaction potentials constructed from the
building blocks ${\cal K}_{\,\,\,\,\,\nu }^\mu  = \sqrt {{g^{\mu
\alpha }}{f_{\alpha \nu }}}$ as
\begin{equation}
{{\cal U}_i} = \sum\limits_{y = 1}^i {{{( - 1)}^{y + 1}}\frac{{(i
- 1)!}} {{(i - y)!}}} {{\cal U}_{i - y}}[{{\cal K}^y}].
\end{equation}

Varying the bulk action (\ref{bulk action}) with respect to the
physical metric ($g_{\mu \nu}$) and $U(1)$ gauge potential
($A_{\mu}$), one arrives at the following relation
\begin{eqnarray} \label{variational principle}
\delta {{\cal I}_{\rm{b}}} &=&  - \frac{1}{{16\pi {G_d}}}\int_{\cal M} {{d^d}x\sqrt { - g} [{G_{\mu \nu }} + \Lambda {g_{\mu \nu }} + m_g^2{{\cal X}_{\mu \nu }} - {T_{\mu \nu }}]} \delta {g^{\mu \nu }} \nonumber \\
&&+ \frac{1}{{8\pi {G_d}}}\int_{\partial {\cal M}} {{d^{d - 1}}x\sqrt { - h} } {n^\alpha }{h^{\mu \nu }}\delta {g_{\mu \nu ,\alpha }} \nonumber \\
&&+ \frac{s}{{4\pi {G_d}}}\int_{\cal M} {{d^d}x\sqrt { - g} } {\nabla _\mu }[{{\cal F}^{s - 1}}{F^{\mu \nu }}]\delta {A_\nu } \nonumber \\
&&+ \frac{s}{{4\pi {G_d}}}\int_{\partial {\cal M}} {{d^{d - 1}}x\sqrt { - h} } {( - {\cal F})^{s - 1}}{n_\mu }{F^{\mu \nu }}\delta {A_\nu },
\end{eqnarray}
where $T_{\mu \nu}$ is the stress-energy tensor as
\begin{equation}
{T_{\mu \nu }} = \frac{1}{2}{g_{\mu \nu }}{\cal L}({\cal F}) - 2{F_{\mu \lambda }}{F_\nu }^{\,\lambda }\frac{{\partial {\cal L}({\cal F})}}{{\partial {\cal F}}}
\end{equation}
and ${\cal X}_{\mu \nu}$ is the consequence of varying dRGT
Lagrangian as
\begin{equation}
{{{\cal X}_{\mu \nu }} =  - \sum\limits_{i = 1}^{d - 2} {\frac{{{c_i}}}{2}\Big[ {{{\cal U}_i}{g_{\mu \nu }} + \sum\limits_{y = 1}^i {{{( - 1)}^y}\frac{{i!}}{{(i - y)!}}{{\cal U}_{i - y}}{\cal K}_{\mu \nu }^y} } \Big]} }.
\end{equation}
Hence, the gravitational and the electromagnetic field equations
are obtained as
\begin{equation} \label{gravitational field eq}
{G_{\mu \nu }} + \Lambda {g_{\mu \nu }} + m_g^2{{\cal X}_{\mu \nu }} = \frac{1}{2}{g_{\mu \nu }}{( - {\cal F})^s} + 2s{F_{\mu \lambda }}{F_\nu }^{\,\lambda }{( - {\cal F})^{s - 1}},
\end{equation}
\begin{equation} \label{electromagnetic field eq}
{\nabla _\mu }[{{\cal F}^{s - 1}}{F^{\mu \nu }}] = 0.
\end{equation}
In order to find TBH solutions, we make use of the following
static ansatz for the physical metric $g_{\mu \nu}$ in $d=n+2$
dimensions as
\begin{equation} \label{physical metric}
d{s^2} =  - V(r)d{t^2} + \frac{{d{r^2}}}{{V(r)}} + {r^2}{h_{ij}}d{x_i}d{x_j}, \,\,\,\,\,\,(i,j = 1,2,3,...,n),
\end{equation}
where the line element ${h_{ij}} d{x_i}d{x_j}$ is the metric of
$n$-dimensional (unit) hypersurface with the constant curvature
${d_{1}}{d_{2}k}$ and volume ${\omega _{n}}$ with the following forms
\begin{equation}
{h_{ij}}d{x_i}d{x_j} =\left\{
\begin{array}{cc}
dx_1^2 + \sum\limits_{i = 2}^{d_{2}} {\prod\limits_{j = 1}^{i - 1} {{{\sin }^2}{x_j}dx_i^2} },  & k=1 \\
dx_1^2 + {\sinh ^2}{x_1}\Big( dx_2^2 + \sum\limits_{i = 3}^{d_{2}} {\prod\limits_{j = 2}^{i - 1} {{{\sin }^2}{x_j}dx_i^2} }\Big), & k=-1 \\
\sum\limits_{i = 1}^{d_{2}} {dx_i^2}, & k=0%
\end{array}%
\right.,
\end{equation}
in which $d_{i}=d-i$ (henceforth we will use this notation). For
the reference metric, we employ the following singular ansatz
\cite{Vegh2013,Cai2015}
\begin{equation} \label{reference metric}
{f_{\mu \nu }} = diag\left( {0,0,c_0^2{h_{ij}}} \right),
\end{equation}
that breaks the translational invariance, a typical property of
many condensed matter systems in the nature
\cite{Vegh2013,Baggioli2015PRL,Baggioli2016JHEP}. Using this
ansatz, the self-interaction potentials are explicitly computed in
arbitrary dimensions as
\begin{equation}
{{\cal U}_i} = {\left( {\frac{{{c_0}}}{r}}
\right)^i}\prod\limits_{j = 2}^{i + 1} {{d_j}},
\end{equation}
indicating a total of $(d-2)$ self-interaction potentials exist in
a $d$-dimensional spacetime.

At this stage, we concentrate on the electromagnetic field
equation (\ref{electromagnetic field eq}). Considering such an
equation and using the static ansatz $A_{\mu}=h(r)
\delta_{\mu}^{0}$, the following differential equation is obtained
for the scalar potential $h(r)$
\begin{equation} \label{differential eq}
(2s - 1)r\frac{{{d^2} h(r)}}{{d{r^2}}} + {d_2}\frac{{d
h(r)}}{{dr}} = 0,
\end{equation}
with the following exact solution
\begin{equation}
h(r) =\phi - \Big( {\frac{{2s - 1}}{{d_{1} - 2s}}} \Big)q{r^{ -
\left( {\frac{{d_{1} - 2s }}{{2s - 1}}} \right)}},
\end{equation}
in which two integration constants $q$ and $\phi$ are related to
the total electric charge and potential of spacetime,
respectively. The constant $\phi$ can be found by imposing the
regularity condition at the horizon, i.e., $A_{t}(r=r_+)=0$,
yielding $\phi = \Big( {\frac{{2s - 1}}{{d_{1} - 2s }}}
\Big)q{r_{+}^{ - \left( {\frac{{d_{1} - 2s }}{{2s - 1}}}
\right)}}$. Therefore, the electric potential at infinity with
respect to the horizon is measured as
\begin{equation} \label{potential}
\Phi= {\left. {{A_\mu }{\chi ^\mu }} \right|_{r \to \infty }} -
{\left. {{A_\mu }{\chi ^\mu }} \right|_{r \to {r_ + }}} = \Big(
{\frac{{2s - 1}}{{d_{1} - 2s }}} \Big)q{r_+^{ - \left(
{\frac{{d_{1} - 2s}}{{2s - 1}}} \right)}},
\end{equation}
where $\chi=\partial_{t}$ is the temporal Killing vector. Note
that the range of nonlinearity parameter $s$ is constrained by
demanding a finite value for the electric potential at infinity
($r \to \infty$), yielding $\frac{1}{2} < s < \frac{{d_{1}}}{2}$.
By implementing the causality and unitarity conditions, we already
found that the nonlinearity parameter $s$ must satisfy the range
$s \ge 1$. So, this together with the asymptotic behavior of the
electric potential imply that the acceptable range of $s$ is
\begin{equation}  \label{power bound}
1 \le s < \frac{{d_{1}}}{2}.
\end{equation}
As will be seen, all of the critical phenomena presented in this
paper meet the above condition. In the limit $s \to 1$, $A_{\mu}$
reduces to the gauge potential of the Maxwell case. Besides, the
interesting case of conformally invariant Maxwell source is
obtained via $s=d/4$. These different cases will be separately
studied.

TBH solutions of the gravitational field equations
(\ref{gravitational field eq}) can be obtained by solving its
$rr$-component, i.e.,
\begin{equation} \label{gravittational field equation}
{{d_2}{d_3}\left( {k - V(r)} \right) - {d_2}r\Big(
{\frac{{dV(r)}}{{dr}}} \Big) + \frac{{d_1}{d_2} }{\ell ^2} r^2+
m_g^2\sum\limits_{i = 1}^{d_{2}} {\Big( {c_0^i{c_i}r_ + ^{2 -
i}\prod\limits_{j = 2}^{i + 1} {{d_j}} } \Big)}  - \frac{{{2^s}(2s
- 1){q^{2s}}}}{{{r^{2\left( {sd_{4} + 1} \right)/(2s - 1)}}}} =
0}.
\end{equation}
This differential equation admits the following black hole
solution
\begin{equation}  \label{metric function}
{V(r) = k + \frac{{r^2}}{{\ell}^2} - \frac{m}{{{r^{{d_3}}}}} +
m_g^2\sum\limits_{i = 1}^{d_{2}} {\Big(
{\frac{{c_0^i{c_i}}}{{{d_2}{r^{i - 2}}}}\prod\limits_{j = 2}^i
{{d_j}} } \Big)}  + \frac{{{2^s}{{(2s -
1)}^2}{q^{2s}}}}{{{d_2}(d_{1} - 2s ){r^{2(sd_{4} + 1)/(2s -
1)}}}}},
\end{equation}
in which $m$ is an integration constant related to the finite mass
of black hole. It is straightforward to check that the obtained
solution satisfies other components of Eq. (\ref{gravitational
field eq}) as well. If we take the massless graviton limit ($m_g
\to 0$), the above metric function reduces to the case of
Schwarzschild-AdS-PMI black hole spacetime reported in Refs.
\cite{HassaineMartinez2007,HassaineMartinez2008,HendiVahidinia2013}.
Also, the charged black hole solution of Maxwell-massive gravity
\cite{CQG2020} is recovered in the limit $s \to 1$. For the
solution obtained, the existence of an essential singularity at
the origin is simply confirmed by examining the behavior of the
curvature scalars such as the Kretschmann scalar, which diverges
only at $r=0$. This singularity is covered by the event horizon
($r_+$), i.e., the largest root of metric function (\ref{metric
function}) with a positive slope. Using the Euclidean trick
\cite{GibbonsHawking1977,HawkingPage1983}, the Hawking temperature
of this spacetime is given by
\begin{equation} \label{temperature-can}
{\beta ^{ - 1}} = T = {\left. {\frac{1}{{4\pi }}\frac{{dV(r)}}
{{dr}}} \right|_{r = {r_ + }}} = \frac{1}{{4\pi {d_2}{r_ +
}}}\Bigg[ {{d_2}{d_3}k +\frac{{d_1}{d_2}}{\ell ^2} {r_{+}^2} +
m_g^2\sum\limits_{i = 1}^{d_{2}} {\Big( {c_0^i{c_i}r_ + ^{2 -
i}\prod\limits_{j = 2}^{i + 1} {{d_j}} } \Big)}  - \frac{{{2^s}(2s
- 1){q^{2s}}}}{{{r_{+}^{2\left( {sd_{4} + 1} \right)/(2s - 1)}}}}}
\Bigg],
\end{equation}
which is in agreement with the definition of surface gravity
\cite{Hawking1975,BardeenCarterHawking1973}. In addition,
following Ref. \cite{Cai2015}, the ADM mass can be obtained
through the Hamiltonian approach as
\begin{equation} \label{ADMmass}
M = \frac{{{d_2}{\omega _n}}}{{16\pi }}m,
\end{equation}
where $m$ is determined from $V(r_+)=0$ as
\begin{equation} \label{constant m}
m = kr_ + ^{{d_3}} + \frac{{r_ + ^{{d_1}}}}{{{\ell ^2}}} +
m_g^2\sum\limits_{i = 1}^{d_{2}} {\Big(
{\frac{{c_0^i{c_i}}}{{{d_2}}}r_ + ^{d_{1} - i}\prod\limits_{j =
2}^i {{d_j}} } \Big) + \frac{{{2^s}{{(2s -
1)}^2}{q^{2s}}}}{{{d_2}(d_{1} - 2s )}}r_ + ^{ - \left(
{\frac{{d_{1} - 2s }}{{2s - 1}}} \right)}}.
\end{equation}

It is worth mentioning that although the asymptotic symmetry group
of black hole solution is not necessarily that of pure AdS, the
validity of relation (\ref{ADMmass}) confirms that the
Ashtekar-Magnon-Das (ADM) mass formula
\cite{Ashtekar1984AMD,Ashtekar2000AMDmass} still holds for black
holes solutions of dRGT massive gravity. In the next section, we
explicitly confirm the above mass formula by evaluating the black
hole partition function and extract all the thermodynamic
quantities.

\section{Semi-Classical Black Hole Partition Function} \label{secIII}

It is believed that all thermodynamic information of a given black
hole system encodes in the associated semi-classical partition
function. Thus, evaluating the black hole's partition function may
be the starting point for studying the BHC at the critical point.
In this section, we are going to evaluate the semi-classical
partition function of the obtained TBHs in the GCE via the
Euclidean path integral formalism \cite{GibbonsHawking1977}.

According to Eq. (\ref{potential}), the Hawking temperature
(\ref{temperature-can}) can be rewritten in terms of fixed
potential as
\begin{equation} \label{temperature-grand}
\beta^{-1}=T = \frac{1}{{4\pi {d_2}{r_ + }}}\Bigg[ {{d_2}{d_3}k +
\frac{{d_1}{d_2}}{\ell ^2} {r_{+}^2}  + m_g^2\sum\limits_{i =
1}^{d_{2}} {\Big( {c_0^i{c_i}r_ + ^{2 - i}\prod\limits_{j = 2}^{i
+ 1} {{d_j}} } \Big)}  - \frac{{{2^s}{{(d_{1} - 2s)}^{2s}}{\Phi
^{2s}}}}{{{{(2s - 1)}^{2s - 1}}r_ + ^{2(s - 1)}}}} \Bigg].
\end{equation}
The partition function in the GCE can be defined by a Euclidean
path integral over the tensor field $g_{\mu \nu}$ and vector gauge
field $A_\mu$ as
\begin{equation} \label{grand canonical parition function}
{{\cal Z}_{\rm{GCE}}} = \int {{\cal D}[g, A]} {e^{ - {{\cal
I}_E}[g, A]}}\simeq e^{ - {{\cal I}_{on-shell}}(\beta,r_+,\Phi)},
\end{equation}
where $\cal D$ denotes integration over all paths and ${\cal I}_E$
represents the Euclidean version of the Lorentzian action ${\cal
I}_G$ by implementing the Wick rotation, ${t_E} = it$
\cite{KlauberBook,Natsuume2015-book,Zee2010-book}. In the GCE, the
electric charge ($Q$) fluctuates, but the associated potential
($\Phi$) is fixed at infinity \cite{York1990GrandCan}. Using the
boundary condition ${\left. {\delta {A_\nu }} \right|_{\partial
{\cal M}}} = 0$ which removes the last surface term in Eq.
(\ref{variational principle}), the electric potential is fixed at
infinity. It leads to the fixed potential boundary condition
necessary for the GCE. In what follows, we implement the
Hawking-Witten prescription (the so-called subtraction method
\cite{HawkingPage1983,Witten1998b}) to evaluate the finite
on-shell action. To do this, we only need to evaluate the on-shell
bulk action (\ref{bulk action}) for the TBH configurations and the
associated AdS backgrounds. It should be noted that these
background solutions are given by setting $m =Q= 0$ in Eq.
(\ref{metric function}), i.e.,
\begin{equation} \label{background}
d{s^2} =  V_0(r)d{t_E^2} + \frac{{d{r^2}}}{ V_0(r)} +
{r^2}{h_{ij}}d{x_i}d{x_j}, \, V_0(r)=k + \frac{{ {r^2}}}{{\ell
^2}}  + m_g^2\sum\limits_{i = 1}^{d_{2}} {\Big(
{\frac{{c_0^i{c_i}}}{{{d_2}{r^{i - 2}}}}\prod\limits_{j = 2}^i
{{d_j}} } \Big)}
\end{equation}
which are not pure AdS in usual sense; it will later become
evident that this choice is the appropriate background, as proved
for massive gravity theories in Refs. \cite{PRD2020,CQG2020}.

In order to find the finite on-shell action, we need to simplify
the form of bulk Lagrangian. Considering the gravitational field
equations (\ref{gravitational field eq}), the Ricci scalar is
obtained as
\begin{equation}
R = \frac{1}{{{d_2}}}\left( {-\frac{ {d_1}{d_2}d}{\ell ^2} + 2m_g^2\chi  - (d - 4s){{( - {\cal F})}^s}} \right),
\end{equation}
where $\chi  \equiv {g^{\mu \nu }}{\chi _{\mu \nu }}$. Regarding
the above relation, the bulk Lagrangian in the action (\ref{bulk
action}) may be rewritten as
\begin{equation}
{{\cal L}_{bulk}} = R +\frac{{d_1}{d_2}}{\ell ^2} +
m_g^2\sum\limits_{i = 1}^{d_{2}} {{c_i}{{\cal U}_i}(g,f) + {{( -
{\cal F})}^s}}  = \frac{2}{{{d_2}}}\Big( {-\frac{{d_1}{d_2}}{\ell
^2}  + m_g^2\sum\limits_{i = 1}^{d_{2}} {(i -
2)\frac{{c_0^i{c_i}}}{{{r^i}}}} \prod\limits_{j = 3}^{i + 1}
{{d_j} + (2s - 1){{( - {\cal F})}^s}} } \Big).
\end{equation}
After tediously long calculation, the on-shell action for the
obtained TBHs is computed as follows
\begin{equation} \label{BH action}
{{\cal I}_{BH}} = \frac{{\beta {\omega _n}}}{{16\pi {G_d}}}\Big[
{\frac{2} {{{\ell ^2}}}{r^{{d_1}}} - m_g^2\sum\limits_{i =
1}^{d_{2}} {\frac{{(i - 2)c_0^i{c_i}}}{{d_{1} - i}}{r^{d_{1} -
i}}\prod\limits_{j = 3}^{i + 1} {{d_j}}  - \frac{{{2^{s +
1}}{{(d_{1} - 2s)}^{2s - 1}}}}{{{d_2}{{(2s - 1)}^{2(s - 1)}}}}r
^{d_{1} - 2s }{\Phi ^{2s}}} } \Big]_{{r_ + }}^R,
\end{equation}
in which $\beta$ is presented in Eq. (\ref{temperature-grand}) and
"$R$" is an upper cutoff regularizing the on-shell action. Doing
the same computation for the case of AdS thermal background, we
arrive at the following relation
\begin{equation} \label{AdS action}
{{\cal I}_{AdS}} = \frac{{\beta_{0} \omega _n}}{{16\pi
{G_d}}}\Big[ {\frac{2} {{{\ell ^2}}}{R^{{d_1}}} -
m_g^2\sum\limits_{i = 1}^{d_{2}} {\frac{{(i -
2)c_0^i{c_i}}}{{d_{1} - i}}{R^{d_{1} - i}}\prod\limits_{j = 3}^{i
+ 1} {{d_j}} } } \Big],
\end{equation}
where $\beta_{0}$ is the period of thermal background. Finally, we
should subtract AdS background action (\ref{AdS action}) from
on-shell (black hole) action (\ref{BH action}). To do so, both the
black hole spacetime and the thermal background must have the same
geometry at $r=R$, i.e., setting ${\beta _0}\sqrt
{{V_0}(r)}|_{r=R} = \beta \sqrt {V(r)}|_{r=R}$ which leads to
${\beta _0} = \beta \left( {1 - \frac{{m{\ell
^2}}}{{2{R^{d_{1}}}}} + O({R^{ - 2(d_{1})}})} \right)$. Now,
applying the mentioned subtraction yields
\begin{eqnarray} \label{on-shell action}
{{\cal I}_{on - shell}} &=& \mathop {\lim }\limits_{R \to \infty } \left( {{{\cal I}_{{\rm{BH}}}} - {{\cal I}_{{\rm{AdS}}}}} \right) \nonumber \\
&=& \frac{{\beta {\omega _n}r_ + ^{{d_3}}}}{{16\pi {G_d}}}\Big[ {k
- \frac{{r_ + ^2}}{{{\ell ^2}}} + m_g^2\sum\limits_{i = 1}^{d_{2}}
{\Big( {\frac{{(i - 1)c_0^i{c_i}}}{{r_ + ^{i - 2}}}\prod\limits_{j
= 3}^i {{d_j}} } \Big) - \frac{{{2^s}{{(d_{1} - 2s)}^{2s -
1}}}}{{{d_2}{{(2s - 1)}^{2(s - 1)}}}}r_ + ^{2(1 - s)}{\Phi ^{2s}}}
} \Big].
\end{eqnarray}

Applying appropriate adjustments, this relation can be examined in
various limits, e.g., one can obtain the finite on-shell action of
Einstein-PMI gravity (as $m_g \to 0$) or that of Maxwell-massive
gravity (as $s \to 1$) or those of Einstein theory or massive
gravity coupled with conformally invariant Maxwell source (as $s
\to d/4$).

\section{Extended Phase Space Thermodynamics} \label{secIV}

This section is devoted to calculating the conserved and
thermodynamic quantities, and examining the first law of
thermodynamics in the extended phase space. First, we can get the
free energy ($F$) from the (finite) on-shell action by using the
relation ${\cal I}_{on - shell}=\beta F$. But, in the extended
phase space ($\Lambda=-\frac{{d_1}{d_2}}{2 \ell^2}=-8 \pi P$), the free energy ($F$) is
identified as the Gibbs free energy (referred to as $G$). Hence,
using Eq. (\ref{on-shell action}), the Gibbs free energy is
computed as
\begin{eqnarray} \label{Gibbs}
{G_\Phi } &=&- {\beta ^{ - 1}}\ln {{\cal Z}_{{\rm{GCE}}}}(T,P,\Phi )  \nonumber \\
&=& \frac{{{\omega _{n}}r_ + ^{{d_3}}}}{{16\pi }}\Big[ {k -
\frac{{16\pi Pr_{+}^2}}{{{d_1}{d_2}}} + m_g^2\sum\limits_{i =
1}^{d_{2}} {\Big( {\frac{{(i - 1)c_0^i{c_i}}}{{r_ + ^{i -
2}}}\prod\limits_{j = 3}^i {{d_j}} } \Big) - \frac{{{2^s}{{(d_{1}
- 2s )}^{2s - 1}}}}{{{d_2}{{(2s - 1)}^{2(s - 1)}}}}r_ + ^{2(1 -
s)}{\Phi ^{2s}}} } \Big].
\end{eqnarray}
Also, the entropy (as an extensive quantity) conjugate to the
temperature (as an intensive quantity) satisfies the so-called
area law ($S = A/4$) in the Einstein and massive gravities (this
can be easily verified by the use of the relation $S =
\frac{1}{4}\int {{d^{d - 2}}x\sqrt {\tilde g} }$, where $\tilde g$
is the induced metric on the horizon). Now, using the Legendre
transformation (${G_\Phi } = H - TS - Q\Phi$), we can find the
corresponding enthalpy ($H$) in the extended phase space. The ADM
mass of TBHs ($M$) which is interpreted as the enthalpy ($H$), can
be obtained in terms of fixed potential ($\Phi$) as follows
\begin{eqnarray} \label{mass-grand}
M &=& {G_\Phi } + TS + \Phi Q  \nonumber  \\
&=& \frac{{{d_2}{\omega _n}}}{{16\pi }}\Big[ {kr_ + ^{{d_3}} +
\frac{{16\pi Pr_ + ^2}}{{{d_1}{d_2}}} + m_g^2\sum\limits_{i =
1}^{d_{2}} {\Big( {\frac{{c_0^i{c_i}}}{{{d_2}}}r_ + ^{d_{1} -
i}\prod\limits_{j = 2}^i {{d_j}} } \Big) + \frac{{{2^s}{{(d_{1} -
2s )}^{2s - 1}}{\Phi ^{2s}}}}{{{d_2}{{(2s - 1)}^{2s - 2}}}}r_ +
^{d_{1} - 2s }} } \Big].
\end{eqnarray}
Considering $P=-({d_1}{d_2})/(16 \pi \ell^2)$ and Eq.
(\ref{potential}), we find that the above relation is in complete
agreement with the obtained ADM mass formula in Eqs.
(\ref{ADMmass}) and (\ref{constant m}). The enthalpy is a function
of $S$ (through the dependency of $r_+$ and $S$), $P$ and $\Phi$,
and therefore, the other thermodynamic quantities can be simply
extracted. The thermodynamic volume and $U(1)$ charge of TBHs are
respectively given by
\begin{equation}
V = {\left( {\frac{{\partial {M}}}{{\partial P}}} \right)_{S,\Phi }} = \frac{{{\omega _{n}}}}{{{d_1}}}r_ + ^{{d_1}},
\end{equation}
and
\begin{equation} \label{charge}
Q = {\left( {\frac{{\partial {M}}}{{\partial \Phi }}} \right)_{S,P}} = \frac{{{\omega _n}{2^{s - 1}}s}}{{4\pi }}{q^{2s - 1}}.
\end{equation}
It is easy to check out the finite electric charge of TBHs is in
line with the result of the usual method of calculating the flux
of the electric field at infinity. In addition, from the enthalpy
(ADM mass) formula (\ref{mass-grand}), the validity of Hawking
temperature in the GCE (\ref{temperature-grand}) is verified
through the use of the standard thermodynamic relation $T =
{\left( {{{\partial M}}/{{\partial S}}} \right)_{P,\Phi}}$.

To sum up, we deduce that all intensive ($P$, $\Phi$, $T$) and
corresponding extensive ($V$, $Q$, $S$) quantities satisfy the
extended first law of thermodynamics in the enthalpy
representation with the following form
\begin{eqnarray} \label{first law}
dM &=& {\left( {\frac{{\partial M}}{{\partial S}}} \right)_{P,\Phi }}dS + {\left( {\frac{{\partial M}}{{\partial P}}} \right)_{S,\Phi }}dP + {\left( {\frac{{\partial M}}{{\partial \Phi }}} \right)_{S,P}}d\Phi \nonumber \\
&=&TdS+VdP+Q d\Phi.
\end{eqnarray}
One can naturally implement the Legendre transform (${G_\Phi } = M
- TS - Q\Phi$) to write down the extended first law in the Gibbs
energy (grand potential) representation as $d{G_\Phi } = -SdT+VdP
-Q d \Phi$.

At this stage, we can discuss the Smarr formula for these types of
TBHs according to the scaling argument. It is confirmed that these
thermodynamic quantities obey the Smarr formula, which for this
class of solutions is given by
\begin{equation} \label{Smarr}
d_{3} M = d_{2} TS - 2PV + \sum\limits_{i = 1}^{d_{2}} {(i - 2)
{{\cal C}_i}{c_i}}  + \frac{{(sd_{4} + 1)}}{{s(2s - 1)}}\Phi Q,
\end{equation}
where
\begin{equation}
{{\cal C}_i} = {\left( {\frac{{\partial M}}{{\partial {c_i}}}}
\right)_{S,P,Q,{c_{j \ne i}}}} = \frac{{{\omega _{{n}}}}}{{16\pi
}}m_g^2c_0^ir_ + ^{{d_{i + 1}}}\prod\limits_{j = 2}^i {{d_j}}.
\end{equation}
The variable ${\cal C}_i$ is a quantity conjugate to the massive
couplings ($c_i$), which is necessary for the consistency of both
the extended first law of thermodynamics and the corresponding
Smarr formula. Note that the pair ${{\cal C}_2}{c_2}$ does not
appear in the Smarr relation (since $c_2$ has zero scaling
\cite{Kastor2009CQG,Hendi2017Mann-PRD}), so one cannot define a
conjugate potential ${\cal C}_2$ in the first law. To conclude,
Eq. (\ref{Smarr}) suggests the extended first law should be
written as
\begin{equation}
dM = TdS + VdP + Qd\Phi  + \sum\limits_{\scriptstyle i = 1\hfill\atop
    \scriptstyle i \ne 2\hfill}^{d_{2} } {{{\cal C}_i}d{c_i}}.
\end{equation}
However, based on AdS/CFT duality, the holographic interpretation
of these new pairs (${{\cal C}_i}{c_i}$) remains an open
question.\footnote{For an example, see Ref.
\cite{Mann2017Sinamuli} for the holographic interpretation of
couplings in the extended phase space thermodynamics of Lovelock
gravity theories.}

\section{Holographic phase transitions} \label{phase transitions} \label{secV}

In order to study BHC at the critical point (or equivalently
holographic phase transitions), we need to examine the equation of
state (EoS) of TBHs. To do so, we first obtain the associated EoS
by inserting $\Lambda=-8 \pi P$ into the Hawking temperature
(\ref{temperature-grand}) as follows
\begin{equation} \label{EOS}
P = \frac{{{d_2}\tilde T}}{{4{r_ + }}} -
\frac{{{d_2}{d_3}k_{\rm{eff}}}} {{16\pi r_ + ^2}} -
\frac{{m_g^2}}{{16\pi }}\sum\limits_{i = 3}^{d_{2}} {\Big(
{\frac{{c_0^i{c_i}}}{{r_ + ^i}}\prod\limits_{j = 2}^{i + 1}
{{d_j}} } \Big)}  + \frac{{{2^s}{{\left( {d_{1} - 2s }
\right)}^{2s}}}}{{16\pi {{\left( {2s - 1} \right)}^{2s -
1}}}}{\left( {\frac{\Phi }{{{r_ + }}}} \right)^{2s}},
\end{equation}
where $k_{\rm{eff}} $ is the effective topological factor \cite{PRD2020}
\begin{equation} \label{eff factor}
k_{\rm{eff}} \equiv [k + m_g^2c_0^2{c_2}]
\end{equation}
and $\tilde T$ is the shifted Hawking temperature (first proposed
in \cite{PVmassive2015PRD})
\begin{equation} \label{shifted Hawking temp}
\tilde T = T - \frac{{m_g^2{c_0}{c_1}}}{{4\pi }}.
\end{equation}
Comparing Eq. (\ref{EOS}) with the expansion of vdW EoS implies
that the event horizon radius (not the thermodynamic volume, $V$)
is associated with the vdW fluid specific volume ($v$) as
\cite{KubiznakMann2012}
\begin{equation}
v = \frac{{4{r_ + }\ell _{\rm{P}}^{{d_2}}}}{{{d_2}}},
\end{equation}
where $\ell _{\rm{P}}$ is the Planck length (here, $\ell _{\rm{P}}
=1$). The possible critical point(s) can be obtained by finding
the inflection point(s) of isotherms in the $P-v$ (or $P-r_+$)
diagrams, i.e.,
\begin{eqnarray} \label{critical points}
{\left( {\frac{{\partial P}}{{\partial v}}} \right)_T} = 0\,\,\, &\leftrightarrow& \,\,\,{\left( {\frac{{\partial P}}{{\partial {r_ + }}}} \right)_T} = 0 \nonumber \\
{\left( {\frac{{{\partial ^2}P}}{{\partial {v^2}}}} \right)_T} = 0\,\,\, &\leftrightarrow& \,\,\,{\left( {\frac{{{\partial ^2}P}}{{\partial {r_+ ^2}}}} \right)_T} = 0.
\end{eqnarray}
The above criteria lead to the following equation for the possible
critical horizon radius (radii)
\begin{equation} \label{critical point equation}
{2{d_2}{d_3}k_{\rm{eff}}r_ + ^{{d_4}} + m_g^2\sum\limits_{i =
3}^{d_{2}} {\Big( {i(i - 1)c_0^i{c_i}r_ + ^{d_{2} -
i}\prod\limits_{j = 2}^{i + 1} {{d_j}} } \Big)}  - \frac{{{2^{s +
1}}s{{(d_{1} - 2s )}^{2s}}{\Phi ^{2s}}}}{{{{(2s - 1)}^{2(s -
1)}}}r_ + ^{( 2s-d_{2})}} = 0}.
\end{equation}
In what follows, we investigate the physical solutions of the
above equation and then analyze the associated holographic phase
transitions case by case with details. However, we should note
that our results in the next sections are generic for all types of
TBHs. In fact, the effective topological factor $k_{\rm{eff}}$
presented in Eq. (\ref{eff factor}) implies that all kinds of TBHs
can exhibit criticality in the same way provided that the same
value is set for $k_{\rm{eff}}$ while keeping other parameters of
the theory fixed (this is always possible by varying the massive
coupling $c_2$). Hence, THBs do exhibit the critical behavior and
the associated phase transition exactly in the same way, even with
the same critical points (first indicated in Ref. \cite{PRD2020}).

It should be noted that the factor  $k_{\rm{eff}}$ is introduced
for both canonical and grand canonical ensembles in PMI-massive
gravity. However, in the context of Maxwell-massive gravity, the
effective factor in the GCE is given by $k_{\rm{eff}} \equiv [k +
m_g^2c_0^2{c_2}-2({d_3}/{d_2})\Phi^2]$, which means that the
$U(1)$ potential ($\Phi$) is absorbed into the effective
topological factor \cite{CQG2020}. The situation is different when
the Maxwell theory is generalized to the PMI theory. Indeed, the
nonlinearity parameter ($s$) does not permit, in general, the
electrostatic potential sector to be absorbed into $k_{\rm{eff}}$.
So, the effective topological factor in the GCE is the same as the
canonical ensemble in PMI-massive gravity ($s \neq 1$).

\subsection{vdW behavior and SBH/LBH phase transition}

The standard vdW behavior takes place when EoS admits only one
physical critical point \cite{KubiznakMann2012}. Evidently,
considering the EoS (\ref{EOS}) up to two interaction potentials
$O({\cal U}_2)$, a critical point may exist if $s \neq 1$. So,
there is a possible critical point given by the following critical
radius in $d \ge 4$ dimensions
\begin{equation}
{r_c} = {\left( {\frac{{{2^s}s{{(d_{1} - 2s )}^{2s}}{\Phi
^{2s}}}}{{{d_2}{d_3}{{(2s - 1)}^{2s - 2}}{k_{{\rm{eff}}}}}}}
\right)^{1/(2s - 2)}},
\end{equation}
in which $k_{{\rm{eff}}}>0$. Obviously, in the massless limit
($m_g=0 \rightarrow k_{\rm{eff}}=k$), the vdW phase transition
takes place only for spherically symmetric AdS black holes
($k=1$). But, the graviton's mass generates additional degrees of
freedom which may lead to the existence of criticality for all
types of TBHs completely in the same way.

\begin{figure}[!htbp]
    $%
    \begin{array}{ccc}
    \epsfxsize=5.1cm \epsffile{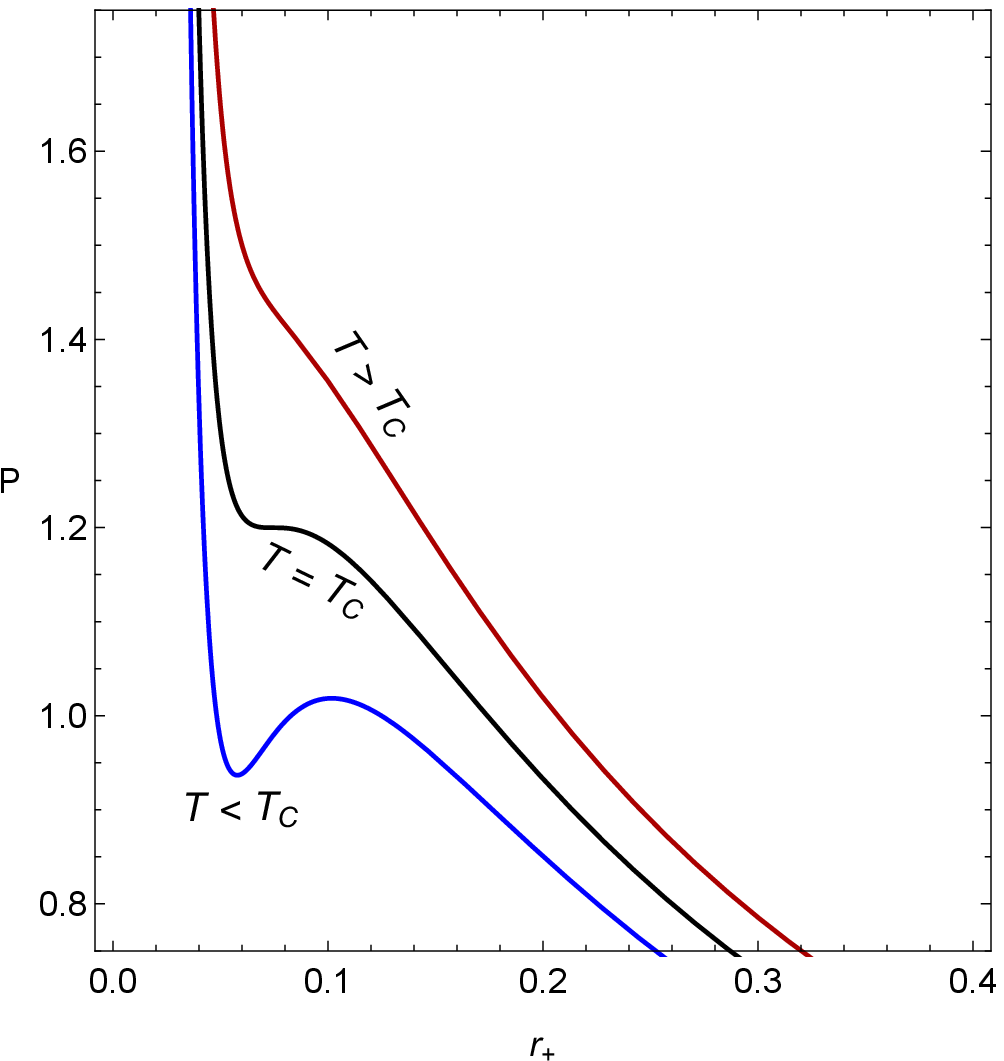}
    \epsfxsize=5.3cm \epsffile{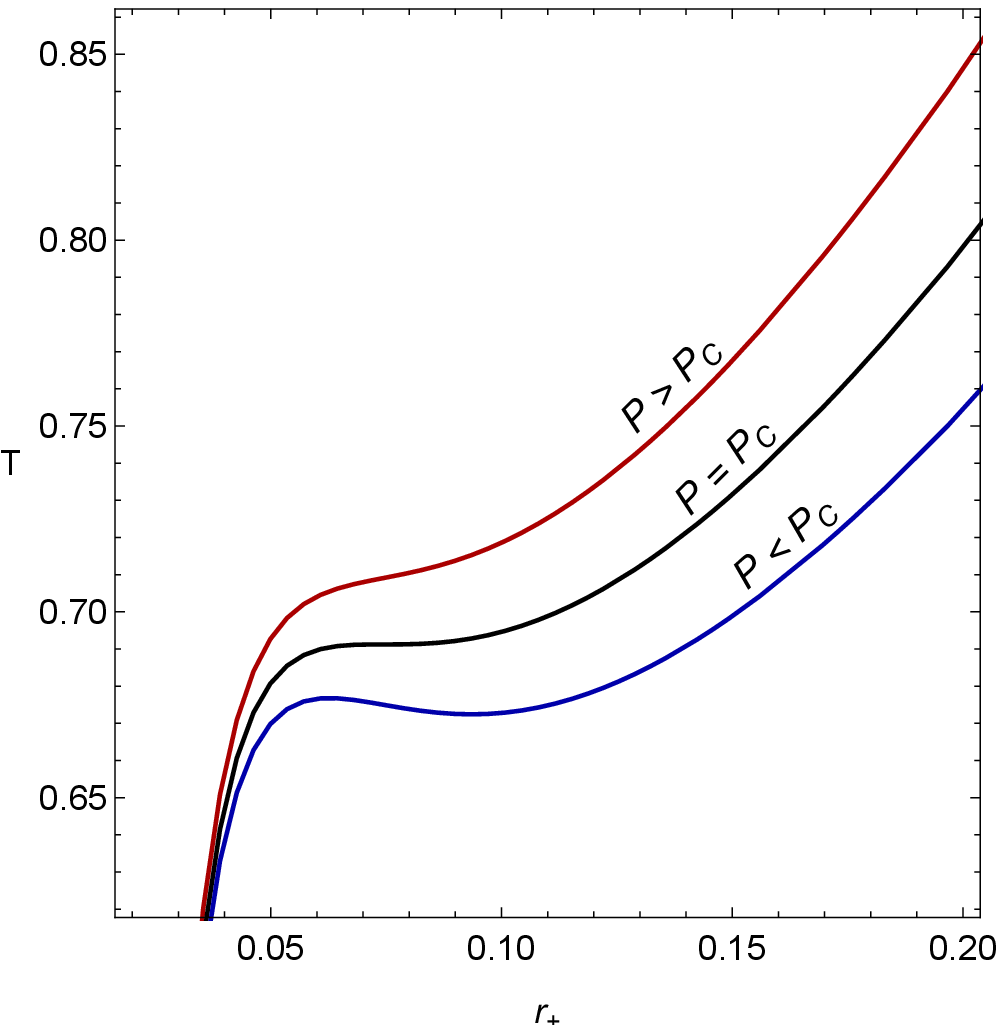}
    \epsfxsize=5.9cm \epsffile{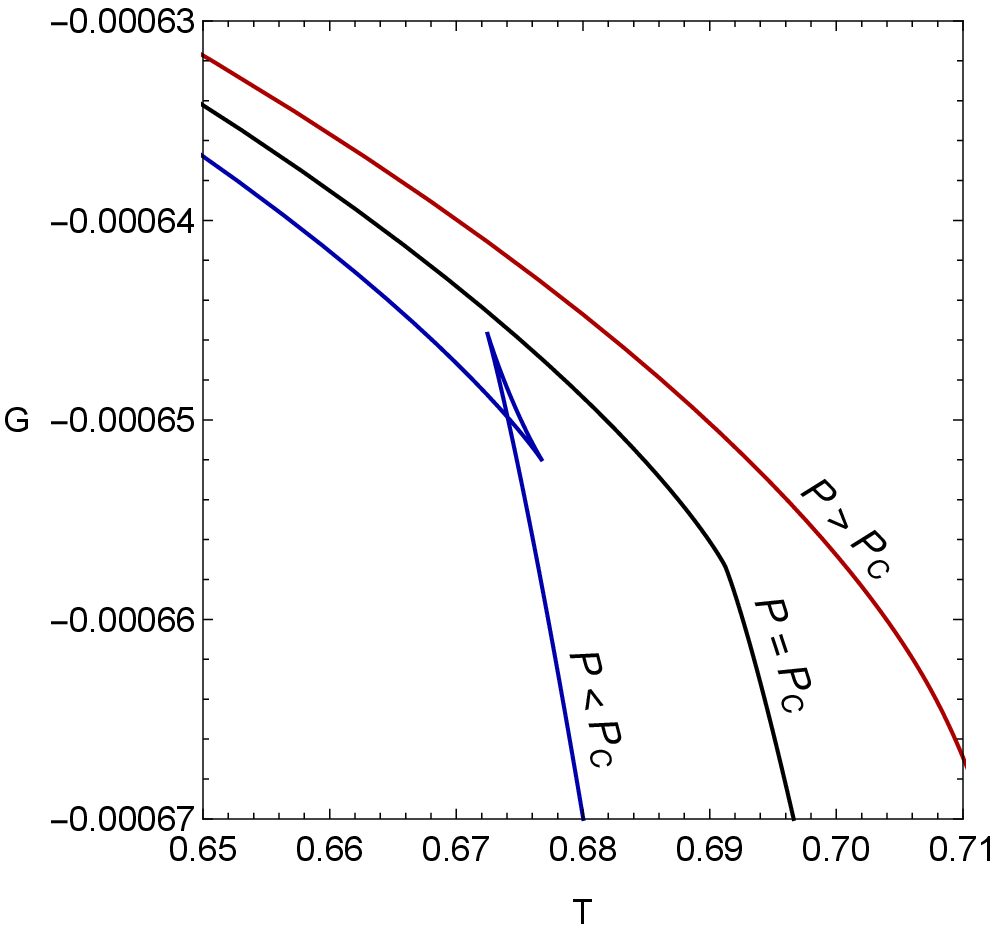}
    &  &
    \end{array}
    $%
    \caption{\textbf{vdW phase transition for PMI-massive TBHs in the GCE:} $P-r_{+}$ (left), $T-r_{+}$
        (middle) and $G-T$ (right) diagrams; we have set $k_{\rm{eff}}=1$, $d=4$,  $m_g=1$, $c=c_{1}=1$, $\Phi =1$ and $s=1.2$ \newline
        \textit{Critical data}: ($T_C=0.6912$, $P_C=1.1998$ and $r_c=0.07434$).}
    \label{PV-vdW}
\end{figure}

To confirm that phase transition takes place with this critical
radius for all types of TBHs, we have depicted Fig. \ref{PV-vdW}.
In the left panel of Fig. \ref{PV-vdW}, the characteristic
behavior of pressure as a function of the event horizon radius is
displayed. As it is observed, for isotherms in the range of
$T<T_C$, an (unphysical) oscillatory part exists which implies the
two phase-behavior (this oscillatory part is replaced by a line of
constant pressure according to Maxwell's equal-area law). For
isotherms with $T>T_C$, the behavior of ideal gas is detected. In
addition, the subcritical isobars of $T-r_+$ diagrams (i.e.,
isobars with $P<P_C$) support the two phase-behavior in agreement
with the result of  $P-r_+$ diagrams, suggesting that the
information of phase transition is encoded in the $T-r_+$ diagrams
as well (for more details on this issue see Ref. \cite{Wei2012}).
Furthermore, the $G-T$ diagram in the right panel of Fig.
\ref{PV-vdW} confirms the characteristic swallowtail behavior of
vdW (first-order) phase transition for isobaric curves with
$P<P_C$, in a way reminiscent of vdW fluid \cite{vanderWaals1873}.
Interestingly, at the critical point, the thermodynamic quantities
up to two interaction potentials $O({\cal U}_2)$ satisfy the
following universal ratio
\begin{equation}
{\rho _c} = \frac{{{P_C}{v_c}}}{{{{\tilde T}_C}}} = \frac{{2s - 1}}{{4s}},
\end{equation}
in terms of the shifted Hawking temperature (\ref{shifted Hawking
temp}). Thus, respecting the constraint on the nonlinearity
parameter Eq. (\ref{power bound}), we always can recover the
well-known vdW ratio as ${\rho_c}=3/8$ if we set $s=2$ in
arbitrary $d \ge 6$ dimensions.

\subsection{RPT and LBH/SBH/LBH phase transition}

The black hole version of reentrant phase transition (RPT), first
reported in Refs. \cite{Mann2012Altamirano,Mann2012Gunasekaran},
can be observed in a black hole configuration whose EoS possesses
two possible critical points, i.e., satisfying criteria in Eq.
(\ref{critical points}). However, the pressure and the temperature
associated with these critical points are positive definite, but,
one of them cannot minimize the Gibbs free energy and
consequently, the global thermal stability fails at this point.
So, only one (physical and second order) critical point $(T_C,
P_C)$ which satisfies criteria (\ref{critical points}) remains in
the phase space. Further investigation shows that
\cite{Mann2012Altamirano} for a certain range of pressure, three
separate phases of black holes (small, intermediate and large)
emerge, indicating the existence of two new critical points in the
phase space. These points are usually referred to as $(T_Z, P_Z)$
and $(T_{\rm {Tr}}, P_{\rm{Tr}})$. At the virtual triple point
$(T_{\rm {Tr}}, P_{\rm{Tr}})$, the first-order and the
zeroth-order coexistence lines join together. Both the first-order
and zeroth-order coexistence lines are terminated at critical
points $(T_C, P_C)$ and $(T_Z, P_Z)$, respectively.

\begin{figure}[!htbp]
    $%
    \begin{array}{ccc}
    \epsfxsize=5.5cm \epsffile{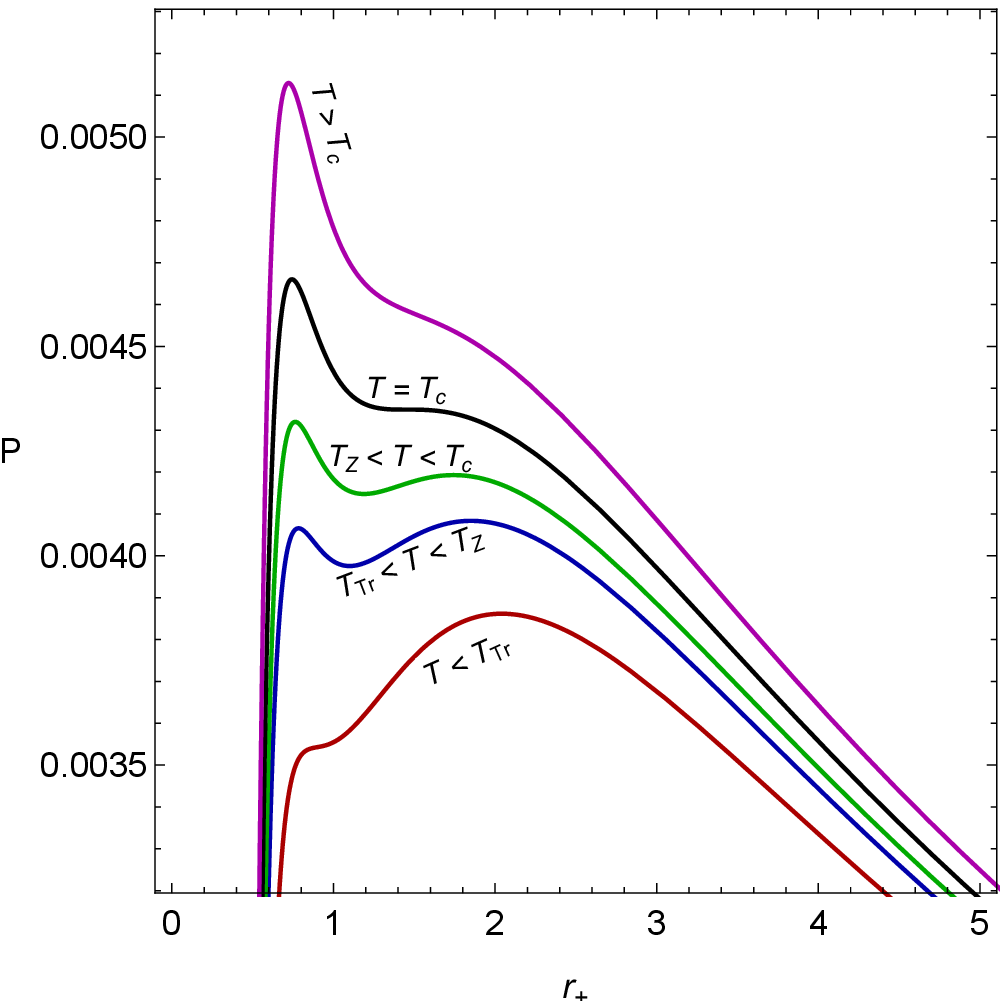}
    \epsfxsize=5.6cm \epsffile{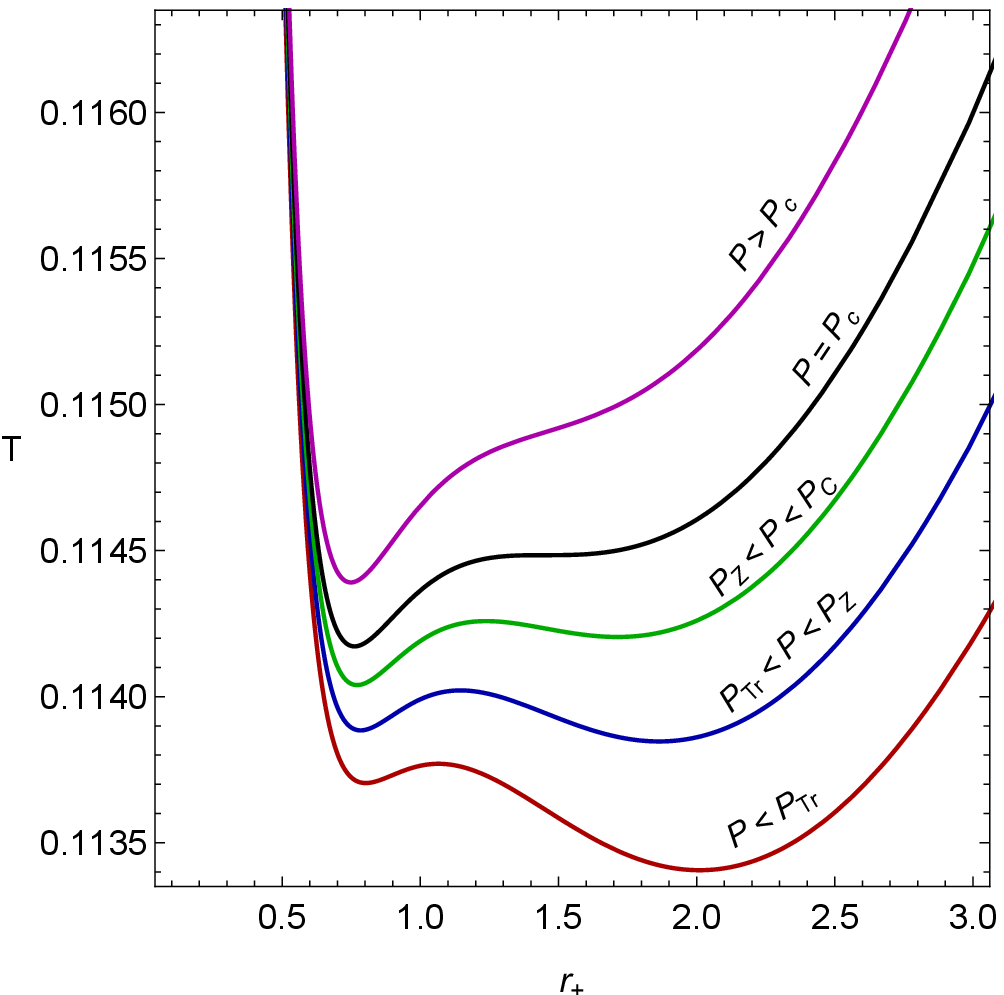}
    \epsfxsize=5.7cm \epsffile{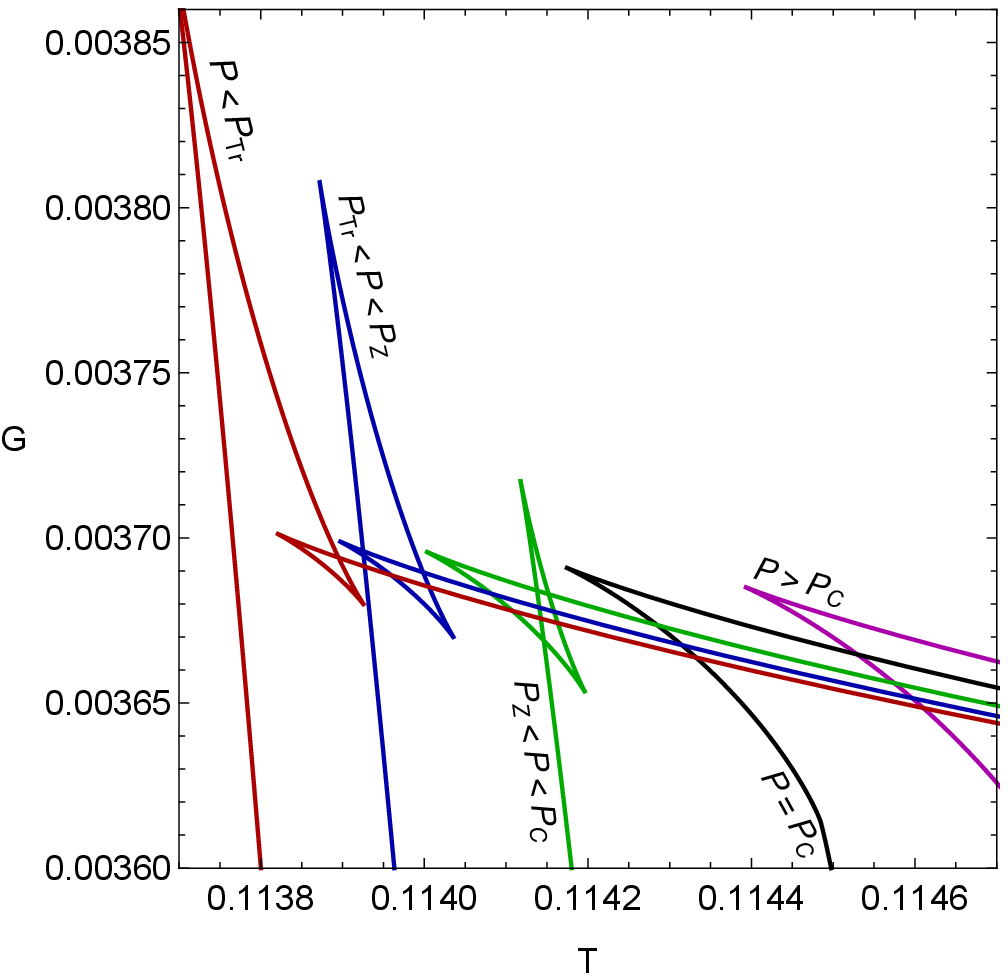}
    &  &
    \end{array}
    $%
    \caption{\textbf{RPT for PMI-massive TBHs in the GCE:} $P-r_{+}$ (left), $T-r_{+}$
        (middle) and $G-T$ (right) diagrams; we have set $k_{\rm{eff}}=0.7$, $d=5$,  $m_g=1$, $c=c_{1}= 1$, $c_3=0.4$, $\Phi =1.3$ and $s=1.3$ \newline
        \textit{Critical data}: ($T_C=0.1145$, $P_C=0.004349$, $r_c =1.4510$), ($T_Z=0.1139$, $P_Z=0.004097$) and ($T_{Tr}=0.1139$, $P_{Tr}=0.004055$)}
    \label{PV-RPT}
\end{figure}

We examined the mentioned conditions for our TBH setups and also
for the other cases in the literature (for example see
\cite{EPJC2019,CQG2020,PRD2020}) indicating this is typical of any
RPT in AdS black hole physics. In order to have RPT phenomenon in
PMI-massive gravity, we have to consider this model at least up to
three graviton's self-interaction terms $O({\cal U}_3)$ and assume
that the other massive couplings vanish in higher dimensions. This
assumption leads to the following equation governing the critical
points for $d \ge 5$ dimensions
\begin{equation}
{d_2}{d_3}k_{\rm{eff}}{r_ + } + 3{d_2}{d_3}{d_4}m_g^2c_0^3{c_3} -
\frac{{{2^s}s{{(d_{1} - 2s)}^{2s}}{\Phi ^{2s}}}}{{{{(2s - 1)}^{2s
- 2}}r_ + ^{(2s - 3)}}} = 0.
\end{equation}
By appropriately fine tuning the parameters, the above equation
could admit two positive roots, indicating the existence of two
possible critical points. So, we have adjusted them to observe the
reentrant behavior of phase transition illustrating in Fig.
\ref{PV-RPT}. In this figure, the $G-T$ diagram simply verify the
existence of the standard first-order phase transition between
large and small black holes for pressures in the range of
$P_Z<P<P_C$. For $P_{\rm {Tr}} <P < P_Z$, by further decreasing
the temperature, a zeroth-order phase transition between small and
intermediate black holes (illustrated by a finite jump for the
isobar in the $G-T$ diagram) takes place along with the standard
first-order phase transition verifying the reentrance of phase
transition. Hence the resultant holographic interpretation is the
sequence of phase transitions between different black hole phases
as LBH $\rightarrow$ SBH $\rightarrow$ IBH. As disclosed in Ref.
\cite{Mann2012Altamirano}, the analog of this critical behavior
(RPT) is also seen in multicomponent fluids and liquid crystals
\cite{Hudson1904,Narayanan1994, LiquidCrystalsBook}.

\subsection{Triple point and LBH/IBH/LBH phase transition} \label{TripleSect}

One of interesting topics related to the BHC is the existence of
analytic EoS for various phenomena such as RPT or triple point.
However in most cases, the critical points cannot obtain
analytically since the corresponding EoS relations are
mathematically complicated and one has to solve them numerically.
In order to find the triple point phenomenon in our setup, we
should consider this model at least up to four graviton's self
interaction potentials $O({\cal U}_4)$. So, the triple point and
the associated LBH/IBH/LBH phase transition will necessarily show
up in $d \ge 6$ dimensions. Assuming that the rest of massive
couplings vanish in higher dimensions, one finds the following
equation for the critical points
\begin{equation}
{d_2}{d_3}k_{\rm{eff}}r_ + ^2 + 3m_g^2c_0^3{c_3}{d_2}{d_3}{d_4}{r_
+ } + 6m_g^2c_0^4{c_4}{d_2}{d_3}{d_4}{d_5} - \frac{{{2^s}s{{(d_{1}
- 2s )}^{2s}}{\Phi ^{2s}}}}{{{{(2s - 1)}^{2s - 2}}r_ + ^{2(s -
2)}}} = 0
\end{equation}
The above equation could admit at most three positive roots,
indicating the existence of three possible critical points. In
order to have a triple point phenomenon, two of those possible
critical points, referred to as ($T_{C_1}, P_{C_1}$) and
($T_{C_2}, P_{C_2}$), necessarily have to be physical (minimizing
the Gibbs free energy). It turns out that another critical point
is always unphysical which cannot minimize the Gibbs free energy.
We confirm that this is typical behavior of any triple point
phenomenon in AdS black holes.

\begin{figure}[!htbp]
    $%
    \begin{array}{ccc}
    \epsfxsize=5.5cm \epsffile{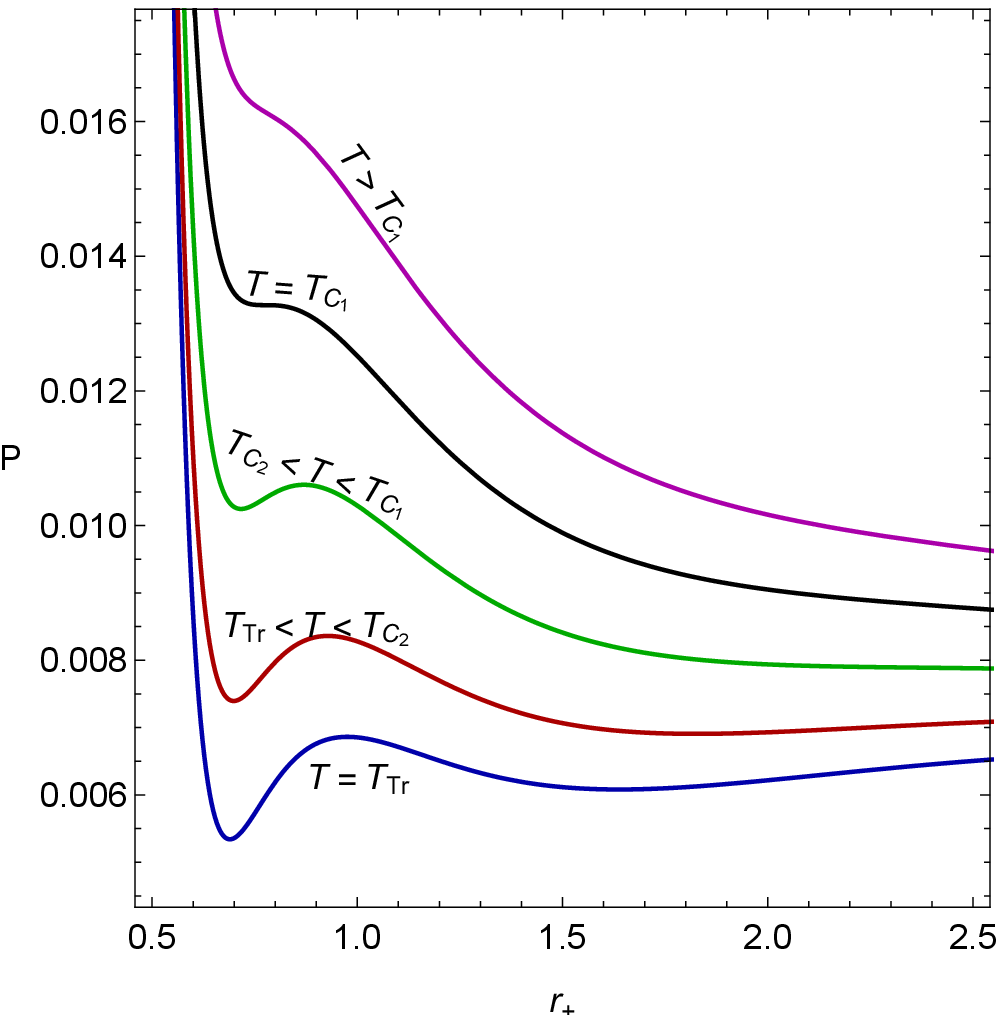}
    \epsfxsize=5.6cm \epsffile{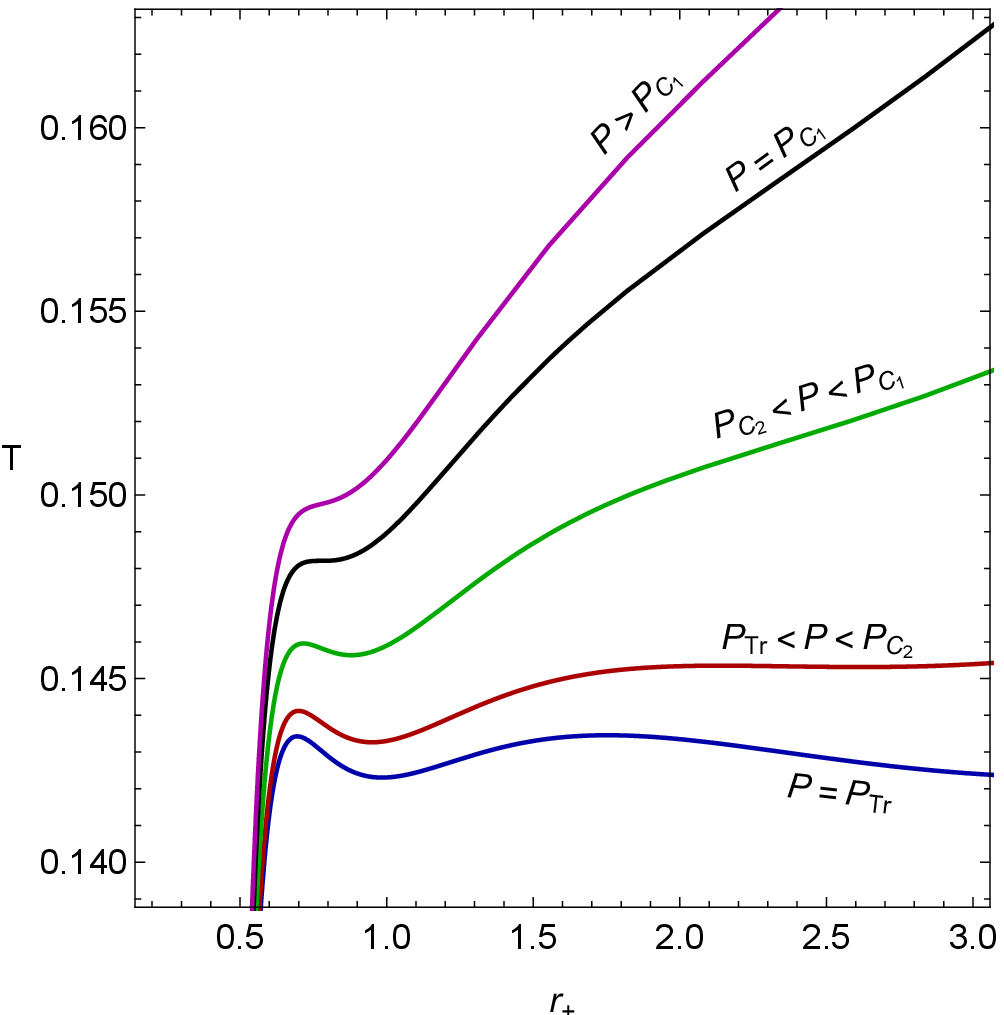}
    \epsfxsize=5.9cm \epsffile{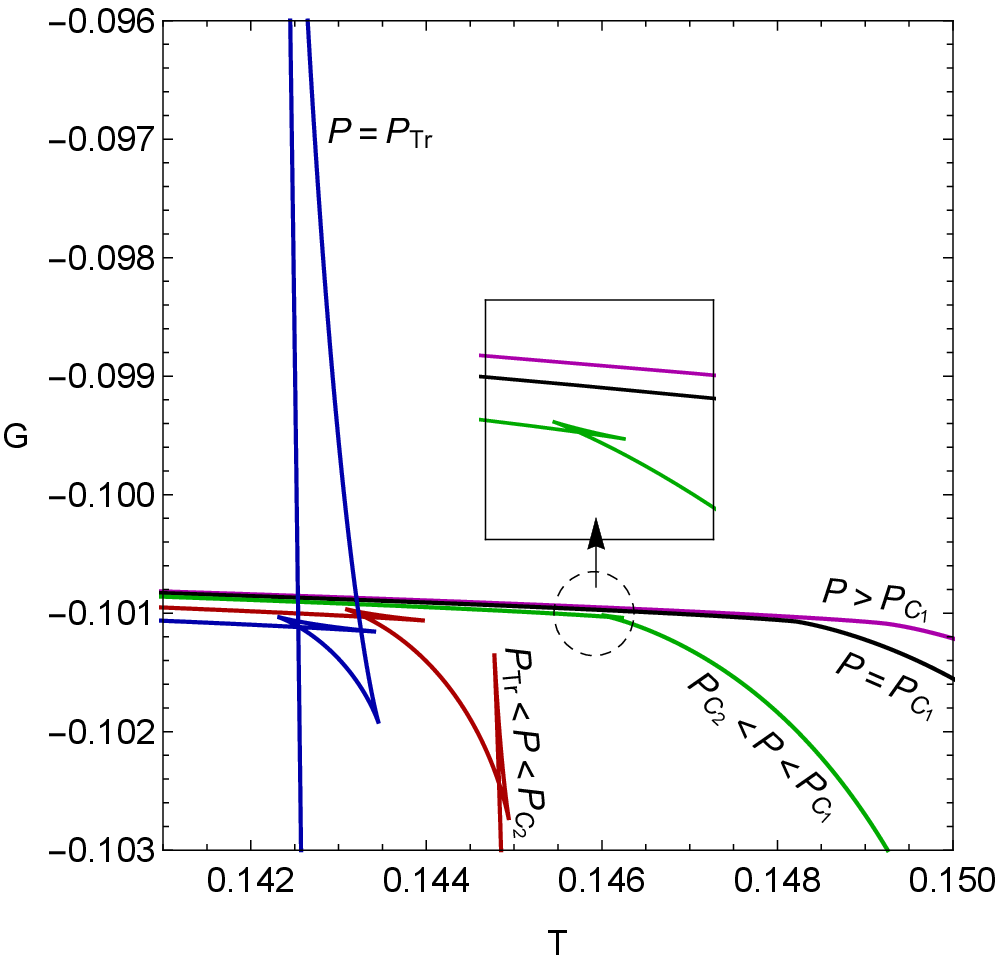}
    &  &
    \end{array}
    $%
    \caption{\textbf{Triple point for PMI-massive TBHs in the GCE:} $P-r_{+}$ (left), $T-r_{+}$
        (middle) and $G-T$ (right) diagrams; we have set $k_{\rm{eff}}=0.9$, $d=6$,  $m_g=1$, $c=c_{1}= 1$, $c_3=-0.7$, $c_4=0.6$, $\Phi =7$ and $s=2.3$ \newline
        \textit{Critical data}: ($T_{C_1}=0.1482$, $P_{C_1}=0.01327$, $r_{c_1}=0.7759$), ($T_{C_2}=0.1456$, $P_{C_2}=0.007732$, $r_{c_2}=2.3588$) and ($T_{Tr}=0.1425$, $P_{Tr}=0.006623$)}
    \label{PV-Triple}
\end{figure}

Solving the above equation, numerically, for a suitable set of
parameters, we have presented an example for this phenomenon,
depicted in Fig. \ref{PV-Triple}. It is obvious that for the
isobars in the range of $P_{C_2}<P<P_{C_1}$, the standard
swallowtail behavior takes place indicating a first-order phase
transition between large and small black holes. By decreasing the
pressure ($P_{Tr}<P<P_{C_2}$), the appearance of two swallowtails
is observed which indicates the emergence of third-phase
(intermediate) and consequently SBH/IBH/LBH phase transition. By
further decreasing the pressure, the mentioned swallowtails merge
at ($T_{Tr}, P_{Tr}$), implying a black hole version of the triple
point. For $P<P_{Tr}$, we only observe the standard vdW behavior
again. In conclusion, the resulting phase structure for TBHs looks
almost like those phase structures of many materials in nature
\cite{Huang2009-book}.

\subsection{Anomalous vdW phase transition}

\begin{figure}[!htbp]
    $%
    \begin{array}{cc}
    \epsfxsize=6.5cm \epsffile{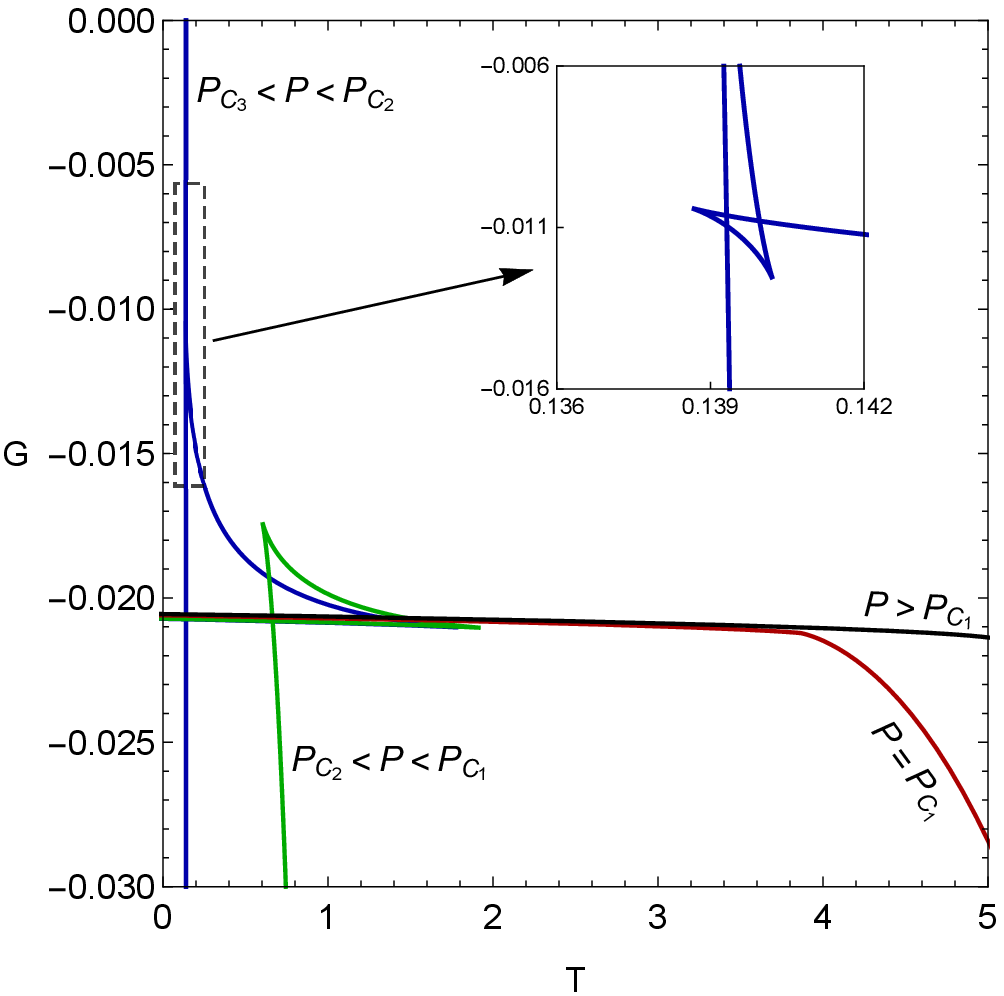}
    \epsfxsize=6.2cm \epsffile{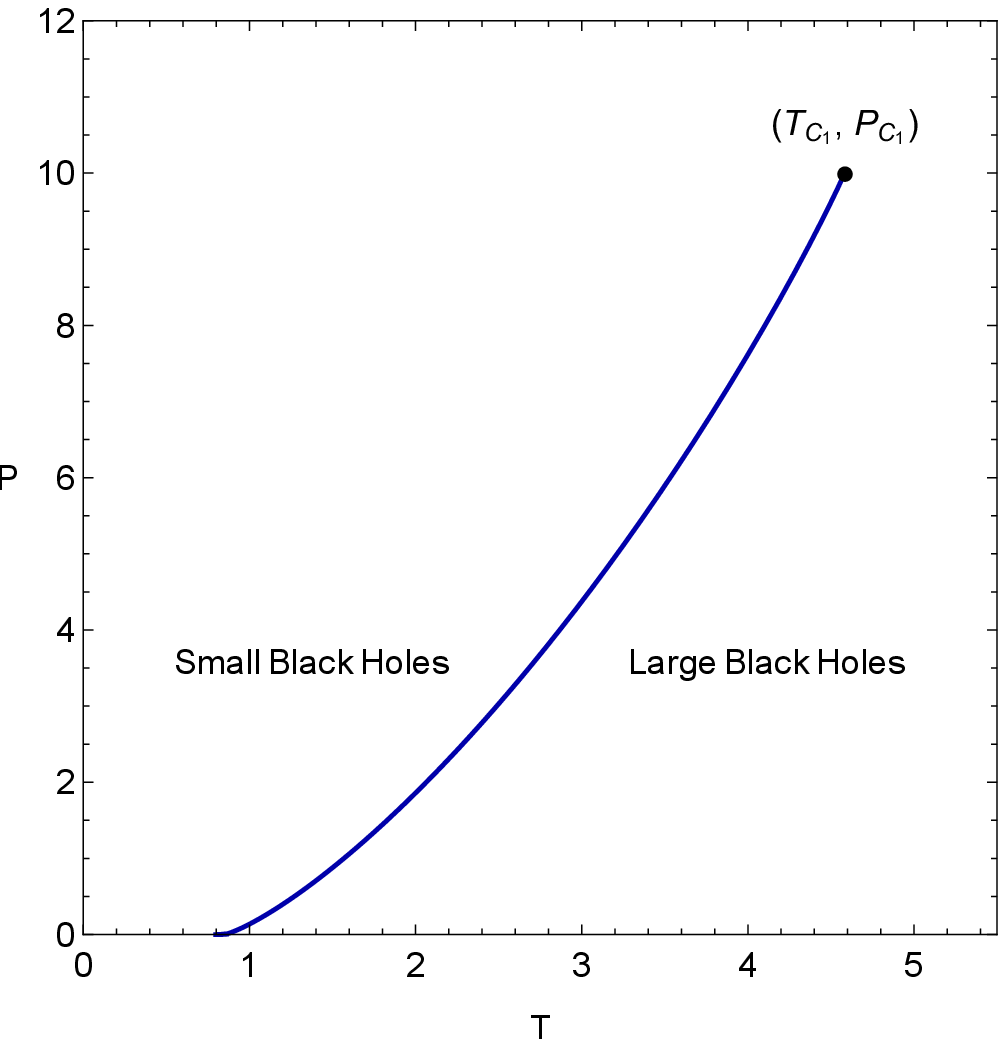}
    &
    \end{array}
    $%
    \caption{\textbf{Anomalous vdW phase transition for PMI-massive TBHs in the GCE:} $G-T$ (left) and $P-T$ (right) diagrams; we have set $k_{\rm{eff}}=0.9$, $d=6$,  $m_g=1$, $c=c_{1}= 1$, $c_3=-0.7$, $c_4=0.6$, $\Phi =4.5$ and $s=2.2$ \newline
        \textit{Critical data}: ($T_{C_1}=3.8665$, $P_{C_1}=10.0013$, $r_{c_1}=0.2409$), ($T_{C_2}=0.1425$, $P_{C_2}=0.006841$, $r_{c_2}=2.6513$) and ($T_{C_3}=0.1346$, $P_{C_3}=0.002542$, $r_{c_3}=1.3965$)}
    \label{PV-vdWlike}
\end{figure}

As conjectured in Ref. \cite{CQG2020}, the anomalous vdW phase
transition (also known as vdW-type phase transition) can
elementally happen by varying the parameter space of the theory in
spacetime dimensions that triple point phenomenon takes place. In
fact, for the case of the triple point, the EoS admits three
possible critical points in which two of them are physical and the
other one is unphysical. So, we can alter at least one of the
parameters in order to change this situation to the case that the
EoS admits three possible (second-order) critical points in which
the only one of them is physical. To do so, here, we vary the
parameters of the nonlinear electromagnetic sector of the previous
example related to the triple point in Sect. \ref{TripleSect} (see
Fig. \ref{PV-Triple}). We choose the new nonlinearity parameter
$s$ and the new potential $\Phi$, respectively, as $s=2.2$ and
$\Phi=4.5$. For this example, the corresponding $G-T$ and $P-T$
diagrams are shown in Fig. \ref{PV-vdWlike}. Evidently, the
resultant holographic phase transition is of SBH/LBH (or vdW) type
with an anomaly in the standard swallowtail behavior. This anomaly
(illustrated in Fig. \ref{PV-vdWlike}) takes place for $P_{C_3} <P
< P_{C_2}$ and does not lead to any new phase transition since it
cannot minimize the Gibbs free energy. In addition, the
corresponding $P-T$ diagram confirms the two-phase behavior,
indicating the well-known SBH/LBH phase transition. Since the
triple point phenomenon is observed in higher dimensions ($d \ge
6$), so the anomalous vdW behavior takes place in $d \ge 6$
dimensions as well.

\section{Comparison with the results of canonical ensemble} \label{sec:comparison}

It is instructive to compare the outcomes of grand canonical
criticality with those counterparts in the canonical (fixed
charge) ensemble. We will use the results of Ref.
\cite{Canonical-PMI-massive}, in which the chemistry of TBHs in
the canonical ensemble was extensively explored within the
framework of PMI-massive gravity. Regarding the canonical
ensemble, it was proved that van der Waals phase transition shows
up in $d \ge 4$, anomalous van der Waals phase transition in $d
\ge 6$ and phase transitions associated with the triple point,
i.e., SBH/IBH/LBH phase transition in $d \ge 6$ dimensions.

Comparing the vdW critical behavior in the canonical and grand
canonical ensembles, we find that the vdW phase transition can
take place for both ensembles in $d\ge4$ dimensions. According to
our numerical investigation we could not find any evidence of a
reentrant behavior in canonical ensemble for TBHs in PMI-massive
gravity. Therefore, as speculated in \cite{Canonical-PMI-massive},
charged TBHs may not exhibit RPT in the canonical ensemble.
Undoubtedly, this speculation relies on numerical investigation
which is not decisive. However, it was shown that this claim can
analytically be proved for $s=1$ in $d = 5$ and $6$ dimensions (by
using numerical investigation in $d = 7, 8, 9$ dimensions, this
phenomenon did not observe too). So the existence of RPT
phenomenon is not ruled out completely in canonical ensemble since
only an analytical proof can judge about this problem. Whether or
not such a phenomenon exist in the canonical ensemble remains an
open question. Besides, we observe that the gravitational version
of triple point and also anomalous vdW behavior can take place in
$d \ge 6$ dimensions for both canonical and grand canonical
ensembles.

One of the interesting results of Refs.
\cite{Canonical-PMI-massive} is that BHC for the case of massive
gravity coupled to both the Maxwell and PMI electrodynamics
yields, qualitatively, the same results in the canonical ensemble.
However, because of nonlinearity parameter $s$, there exist more
possibilities (and consequently more parameter space) for phase
transition in PMI-massive gravity. But, comparing the GCE of TBHs
in Maxwell-massive gravity and PMI-massive gravity models exhibits
that each critical phenomenon commences appearing in diverse
dimensions depending on the gravity model one is dealing with. So,
in PMI-massive gravity, these phenomena can show up in lower
dimensions with respect to Maxwell-massive gravity. We have
summarized these results in Table \ref{tab:ensembles}. In
conclusion, BHC in the GCE in comparison with canonical ensemble
for both models is completely different and seemingly richer.

\begin{table}[]
    \caption{Holographic phase transitions in Einstein and massive gravity theories coupled with nonlinear electromagnetic field
    (i.e., PMI electrodynamics) for the canonical (fixed charge) \cite{Canonical-PMI-massive} and grand canonical (fixed potential) ensembles.}
\label{tab:ensembles}
    \begin{tabular}{cccll}
        \cline{1-3}
        \multicolumn{1}{|c|}{\textbf{Theory}} & \multicolumn{1}{c|}{\textbf{Canonical ensemble}} & \multicolumn{1}{c|}{\textbf{GCE}} &  &  \\ \cline{1-3}
        \multicolumn{1}{|c|}{\textbf{\begin{tabular}[c]{@{}c@{}}Einstein-Maxwell gravity\\ ($s=1$)\end{tabular}}} & \multicolumn{1}{c|}{\begin{tabular}[c]{@{}c@{}}vdW in $d \ge 4$\\ (only spherical BHs) \end{tabular}} & \multicolumn{1}{c|}{nothing} & \multicolumn{1}{r}{{\ul \textbf{}}}  &   \\ \cline{1-3}
        \multicolumn{1}{|c|}{\textbf{\begin{tabular}[c]{@{}c@{}}Einstein-PMI gravity\\ $1/2 < s < d_{1}/2$\end{tabular}}} &\multicolumn{1}{c|}{\begin{tabular}[c]{@{}c@{}}vdW in $d \ge 4$\\ (only spherical BHs) \end{tabular}} & \multicolumn{1}{c|}{\begin{tabular}[c]{@{}c@{}}vdW in $d \ge 4$\\ (only spherical BHs) \end{tabular}} &  &  \\ \cline{1-3}
        \multicolumn{1}{|c|}{\textbf{\begin{tabular}[c]{@{}c@{}}Maxwell-massive gravity\\ ($s=1$)\end{tabular}}} & \multicolumn{1}{c|}{\begin{tabular}[c]{@{}c@{}}vdW in $d \ge 4$\\ Anomalous vdW in $d \ge 6$\\ Triple point in $d \ge 6$\end{tabular}} & \multicolumn{1}{c|}{\begin{tabular}[c]{@{}c@{}}vdW in $d \ge 5$\\ RPT in $d \ge 6$\\ Anomalous vdW in $d \ge 7$\\ Triple point in $d \ge 7$\end{tabular}} & \multicolumn{1}{r}{{\ul \textbf{}}} &  \\ \cline{1-3}
        \multicolumn{1}{|c|}{\textbf{\begin{tabular}[c]{@{}c@{}}PMI-massive gravity\\ $1/2 < s < d_{1}/2$\end{tabular}}} & \multicolumn{1}{c|}{\begin{tabular}[c]{@{}c@{}}vdW in $d \ge 4$\\ Anomalous vdW in $d \ge 6$\\ Triple point in $d \ge 6$\end{tabular}} & \multicolumn{1}{c|}{\begin{tabular}[c]{@{}c@{}}vdW in $d \ge 4$\\ RPT in $d \ge 5$ \\ Anomalous vdW in $d \ge 6$\\ Triple point in $d \ge 6$\end{tabular}} &  &  \\ \cline{1-3}
        \multicolumn{1}{l}{} & \multicolumn{1}{l}{} & \multicolumn{1}{l}{} &  &
    \end{tabular}
\end{table}

\section{Gravity coupled with conformally invariant Maxwell source} \label{sec:Massive-CIM}

In this section, we briefly discuss the case of (massive) gravity
coupled with conformally invariant Maxwell source, i.e., setting
$s=d/4$ for the PMI Lagrangian (\ref{PMI Lagrangian}). The matter
sector consists of a conformally invariant extension of Maxwell
Lagrangian as ${\cal L}_{\rm{CIM}} = (-F_{\mu \nu } F^{\mu
\nu})^{d/4}$ (for more discussions see Refs.
\cite{HassaineMartinez2007,HassaineMartinez2008}) and one can
check out that the black hole solutions exist in this model as
well. The bulk action of this model can be written as
\begin{equation} \label{CIM bulk action}
{{\cal I}_{\rm{b}}} =  - \frac{1}{{16\pi {G_d}}} \int_{\cal M}
{{d^{d}}x\sqrt { - g} \Big[R - 2\Lambda  + m_g^2\sum\limits_{i =
1}^{d_{2}} {{c_i}{{\cal U}_i}(g,f)}  +{(-{F_{\mu \nu }}{F^{\mu \nu
}})^{d/4}} \Big]}.
\end{equation}
It is easy to show that the matter action enjoys conformal
invariance (i.e., ${g_{\mu \nu }} \to {\Omega ^2}{g_{\mu \nu }}$
and ${A_\mu } \to {A_\mu }$) and the associated energy-momentum
tensor is given by
\begin{equation}
{T_{\mu \nu }} = \frac{1}{2}{g_{\mu \nu }}{( - {\cal
F})^{\frac{d}{4}}} + \frac{d}{2}{F_{\mu \lambda }}{F_\nu
}^{\,\lambda }{( - {\cal F})^{\frac{d}{4}-1}},
\end{equation}
which is traceless $T_\mu ^\mu  = 0$. The (static) charged TBH
solutions with a purely radial electric field are easily found by
using our previous line element ansatz in Eq. (\ref{physical
metric}) with the singular (degenerate) reference metric defined
in Eq. (\ref{reference metric}). Also the metric function $V(r)$
is simply obtained by inserting $s=d/4$ into the relation
(\ref{metric function}). So, like the case of Einstein gravity
coupled with the CIM electromagnetic source, TBH solutions appear
for the context of CIM-massive gravity model in arbitrary
dimensions as well. Subsequently, the entire thermodynamic
quantities such as temperature, entropy, electric potential and
charge, ADM mass, volume, etc can be obtained by the use of these
substitutions for the case of (massive) gravity with a conformally
invariance Maxwell source. We do not intend to repeat them here
and just mention that these quantities satisfies the Smarr
relation with the following explicit form
\begin{equation} \label{Smarr-CIM}
d_{3}M = d_{2}TS - 2PV + \sum\limits_{i = 1}^{d_{2}} {(i -
2){{\cal C}_i}{c_i}}  + \frac{2d_{2}}{d}\Phi Q.
\end{equation}
Now, we concentrate on the black hole EoS in both canonical (fixed
charge) and grand canonical (fixed potential) ensembles. EoS
relations in the canonical and grand canonical ensembles are
respectively given by
\begin{equation} \label{pressure-can-CIM}
{P_{can}} = \frac{{{d_2}\tilde T}}{{4{r_ + }}} - \frac{{{d_2}{d_3}
{k_{{\rm{eff}}}}}}{{16\pi r_ + ^2}} - \frac{{m_g^2}}{{16\pi
}}\sum\limits_{i = 3}^{d_{2}} {\Big( {\frac{{c_0^i{c_i}}}{{r_ +
^i}}\prod\limits_{j = 2}^{i + 1} {{d_j}} } \Big)}  +
\frac{{{2^{d/4}}\;d_{2}}}{{32\pi }}{\left( {\frac{{\sqrt q }}{{{r_
+ }}}} \right)^d},
\end{equation}
and
\begin{equation} \label{pressure-grand-CIM}
P_{grand} = \frac{{{d_2}\tilde T}}{{4{r_ + }}} - \frac{{{d_2}{d_3}
{k_{{\rm{eff}}}}}}{{16\pi r_ + ^2}} - \frac{{m_g^2}}{{16\pi
}}\sum\limits_{i = 3}^{d_{2}} {\Big( {\frac{{c_0^i{c_i}}}{{r_ +
^i}}\prod\limits_{j = 2}^{i + 1} {{d_j}} } \Big)}  +
\frac{{{2^{d/4}}\; {d_{2}} }}{{32\pi }}{\left( {\frac{\Phi }{{{r_
+ }}}} \right)^{d/2}}.
\end{equation}
Let us first discuss the case of GCE. It is inferred from Eq.
(\ref{pressure-grand-CIM}) that the final results of the TBH phase
transitions in the fixed potential ensemble are qualitatively the
same as the neutral TBHs of pure massive gravity \cite{PRD2020} or
the charged TBHs in the fixed potential ensemble of
Maxwell-massive gravity \cite{CQG2020} in even dimensions ($d=4,
6, 8, ...$). Indeed, the electromagnetic sector in EoS
(\ref{pressure-grand-CIM}) is absorbed into the $i(=d/2$)th
massive term. For instance, in $d=6$, the vdW and RPT phenomena
are observed, exactly the same as the $P-v$ criticality of
(electrically) neutral TBHs in pure massive gravity \cite{PRD2020}
or charged TBHs in Maxwell-massive gravity in the GCE
\cite{CQG2020}. However, there is no criticality for $d=4$
dimensions, but the vdW phase transition could take place in $d=5$
dimensions. The other critical phenomena, i.e., anomalous vdW and
triple point, will show up in $d \ge 7$ dimensions.

In the case of canonical ensemble, we observe the standard vdW
phase transition in $d \ge 4$ dimensions but this time for all
kinds of TBHs in the context of massive gravity. For $d \ge 6$,
besides the standard vdW, the anomalous vdW and triple point
phenomena are also observed. However, by the numerical
investigation, we could not find evidence for the RPT phenomenon.

In addition, the vanishing of graviton's mass ($m_g \to 0$) leads
to Einstein gravity coupled to a CIM electromagnetic source, and
regarding this model, one can straightforwardly deduce the
following items: the vdW critical behavior occurs for $d \ge 4$
dimensions in the canonical ensemble and for $d \ge 5$ dimensions
in the GCE.  Obviously, criticality takes place only for
spherically symmetric AdS black holes. We have summarized the
results of criticality in table \ref{tab:ensemblesCIM} for
convenience.

\begin{table}[]
    \caption{Holographic phase transitions in Einstein and massive gravity theories with a CIM source for the canonical and the grand
    canonical ensembles.} \label{tab:ensemblesCIM}
    \begin{tabular}{cccll}
        \cline{1-3}
        \multicolumn{1}{|c|}{\textbf{Theory}} & \multicolumn{1}{c|}{\textbf{Canonical ensemble}} & \multicolumn{1}{c|}{\textbf{GCE}} &  &  \\ \cline{1-3}
        \multicolumn{1}{|c|}{\textbf{\begin{tabular}[c]{@{}c@{}}Einstein gravity coupled to \\ CIM source $(s=d/4)$ \end{tabular}}} & \multicolumn{1}{c|}{\begin{tabular}[c]{@{}c@{}}vdW for $d \ge 4$\\ (only spherical BHs) \end{tabular}} &  \multicolumn{1}{c|}{\begin{tabular}[c]{@{}c@{}}vdW for $d \ge 5$\\ (only spherical BHs) \end{tabular}} &  &  \\ \cline{1-3}
        \multicolumn{1}{|c|}{\textbf{\begin{tabular}[c]{@{}c@{}}Massive gravity coupled to \\ CIM source ($s=d/4$)\end{tabular}}} & \multicolumn{1}{c|}{\begin{tabular}[c]{@{}c@{}}vdW for $d \ge 4$\\ Anomalous vdW for $d \ge 6$\\ Triple point for $d \ge 6$\end{tabular}} & \multicolumn{1}{c|}{\begin{tabular}[c]{@{}c@{}}vdW for $d \ge 5$\\ RPT for $d \ge 6$\\ Anomalous vdW for $d \ge 7$\\ Triple point for $d \ge 7$\end{tabular}} &  &  \\ \cline{1-3}

        \multicolumn{1}{l}{} & \multicolumn{1}{l}{} & \multicolumn{1}{l}{} &  &
    \end{tabular}
\end{table}

\section{Critical exponents} \label{sec:exponents}

In this section, we intend to compute the independent critical
exponents ($\alpha$, $\beta$, $\gamma$ and $\delta$)
\cite{Huang2009-book,Kaputsa2006-book,Zinn-Justin1996-book}
associated to the obtained nonlinearly charged TBHs in PMI-massive
gravity. We show that these exponents match the vdW fluid system,
i.e., $(\alpha, \beta, \gamma, \delta)=(0, 1/2, 1, 3)$. These
independent exponents are listed in the following relations
\cite{KubiznakMann2012}
\begin{equation} \label{alpha}
{C_v} = T{\left( {\frac{{\partial S}}{{\partial T}}} \right)_v}
\propto {\left| \tau \right|^{ - \alpha }},
\end{equation}
\begin{equation} \label{beta}
\eta = v_l - v_g \propto |\tau|^{\beta},
\end{equation}
\begin{equation} \label{gamma}
{\kappa _T} =  - \frac{1}{v}{\left. {\frac{{\partial v}}{{\partial P}}} \right|_T} \propto {\left| \tau  \right|^{ - \gamma }},
\end{equation}
\begin{equation} \label{delta}
\left| {P - {P_c}} \right| \propto {\left| {v - {v_c}} \right|^\delta },
\end{equation}
 in terms of the following reduced variables
\begin{equation}
p \equiv \frac{P}{{{P_C}}}, \quad\tau  \equiv \frac{{T -
        {T_C}}}{{{T_C}}}, \quad w \equiv \frac{{v - {v_c}}}{{{v_c}}},
\quad \rho_c \equiv \frac{P_C v_c}{T_C}.
\end{equation}
We perform this analysis for both canonical and grand canonical
ensembles. First of all, we determine the exponent $\alpha$. For
static TBHs, we have $C_v=0$, so the exponent $\alpha$ is simply
equal to zero ($\alpha=0$). Now, by studying the equation of state
near the critical point, we can determine the rest of them.
Expanding the equation of states in the vicinity of critical point
yields
\begin{equation} \label{EOS expansion}
p = 1 + \frac{\tau }{{{\rho _c}}}(1 - w) + h({v_c},{c_i},s){w^3} + O(\tau{w^2},{w^4}),
\end{equation}
where the details of function $h(v_c,c_i,s)$ does not alter the
final result. Schematically, this kind of expansion for any black
hole setup ensures that the critical exponents match the standard
critical exponents of vdW fluid, as will be clarified in a moment.
The explicit form of the function $h(v_c,{c_i},s)$ in the
canonical ensemble is obtained as\footnote{In the limit $s \to 1$,
the function $h(v_c,c_i,s)$ should reduce to the Maxwell-massive
case in Ref. \cite{CQG2020}. For the electromagnetic part of this
function, there is a mismatch between the coefficients of Eq.
\ref{h function can} here (as $s \to 1$) and its counterpart in
Ref. \cite{CQG2020}. Actually, to correctly match the coefficients
we note that there is a minor typo in Ref. \cite{CQG2020}.
However, as mentioned, the details of this function is not
important at all and does not change the final result.}
\begin{eqnarray} \label{h function can}
h &\equiv& \frac{{m_g^2{c_0}{c_1} - 4\pi {T_C}}}{{4\pi {P_C}{v_c}}} + \frac{{4{d_3}{k_{{\rm{eff}}}}}}{{{d_2}\pi {P_C}v_c^2}}- \frac{{{2^{s-9} \times 4^{2(ds - 1)/(2s - 1)}}s(2sd_{1} - 1)(sd - 1){q^{2s}}}}{{{3 }\pi {P_C}{{(s - 1/2)}^2}{d^{(2sd_{3}+ 1)/(2s - 1)}}v_c^{2s{d_2}/(2s - 1)}}}  \nonumber \\
&&+ \frac{{40m_g^2c_0^3{c_3}{d_3}{d_4}}}{{d_2^2\pi {P_C}v_c^3}} + \frac{{320m_g^2c_0^4{c_4}{d_3}{d_4}{d_5}}}{{d_2^3\pi {P_C}v_c^4}} + O\left( {\frac{1}{{v_c^5}}} \right),
\end{eqnarray}
and in the GCE is given by
\begin{eqnarray}
h &\equiv& \frac{{m_g^2{c_0}{c_1} - 4\pi {T_C}}}{{4\pi {P_C}{v_c}}} + \frac{{4{d_3}{k_{{\rm{eff}}}}}}{{{d_2}\pi {P_C}v_c^2}} - \frac{{s(s + 1)({s^2} - 1/4){2^{5s}}}}{{6\pi P_C }}{\left( {\frac{{(d_{1} - 2s)\Phi }}{{(2s - 1){d_2}{v_c}}}} \right)^{2s}} \nonumber \\
&&+ \frac{{40m_g^2c_0^3{c_3}{d_3}{d_4}}}{{d_2^2\pi {P_C}v_c^3}} +
\frac{{320m_g^2c_0^4{c_4}{d_3}{d_4}{d_5}}}{{d_2^3\pi {P_C}v_c^4}}
+ O\left( {\frac{1}{{v_c^5}}} \right).
\end{eqnarray}
To obtain the exponent $\beta$, which determines the behavior of
the order parameter $\eta$ on the isotherms, we have to
differentiate the expansion (\ref{EOS
    expansion}) for a fixed $\tau<0$, yielding
\begin{equation}
dp = \Big( { - \frac{\tau}{{{\rho _c}}} + 3h(v_c,{c_i},s){w^2}} \Big)dw.
\end{equation}
Inserting this into the Maxwell's equal area law ($\oint {vdP = 0}$) leads to
\begin{equation} \label{Maxwell's construction}
\int_{{w_l}}^{{w_s}} {wdp = } \frac{{ - \tau}}{{2{\rho _c}}}(w_s^2 - w_l^2) + \frac{{3h}}{4}(w_s^4 - w_l^4) = 0,
\end{equation}
where ${w_l}$ and ${w_s}$ denote the volume of large and small
black holes, respectively. The pressure of different black hole
phases keeps unchanged at the critical point, and therefore, for
the expansion of EoS (\ref{EOS expansion}) one arrives at
\begin{equation} \label{pressureATtransition}
1 + \frac{\tau}{{{\rho _c}}}(1 - {w_s}) + h(v_c,{c_i},s)w_s^3 = 1 + \frac{\tau}{{{\rho _c}}}(1 - {w_l}) + h(v_c,{c_i},s)w_l^3,
\end{equation}
Considering Eqs. (\ref{Maxwell's construction}) and
(\ref{pressureATtransition}), simultaneously, one can obtain a
nontrivial solution as
\begin{equation}
{w_s} =  - {w_l} = \sqrt {\frac{{ - \tau}}{{{\rho _c}h}}},
\end{equation}
implying
\begin{equation}
\eta = v_l - v_s = 2 v_c  \sqrt {\frac{{ - \tau}}{{{\rho _c}h}}} \propto |\tau|^{1/2}.
\end{equation}
According to Eq. (\ref{beta}), this result confirms that
$\beta=1/2$ for both canonical and grand canonical ensembles.

Now, from the definition of the isothermal compressibility near
the critical point, Eq. (\ref{gamma}), we can find the exponent
$\gamma$. To do this, considering both ensembles, we differentiate
the expansion of EoS (\ref{EOS expansion}) to get
\begin{equation}
{\left. {\frac{{\partial P}}{{\partial v}}} \right|_T} = \frac{{ - {P_c} \tau}}{{{\rho _c}{v_c}}}  + O(\tau w,{w^2}).
\end{equation}
Using ${\left. {\frac{{\partial v}}{{\partial P}}} \right|_T} =
{\left( {{{\left. {\frac{{\partial P}}{{\partial v}}} \right|}_T}}
\right)^{ - 1}}$ and the above relation, it is confirmed that
\begin{equation}
{\kappa _T} =  - \frac{1}{v}{\left. {\frac{{\partial v}}{{\partial P}}} \right|_T} \propto \frac{{{\rho _c}{v_c}}}{{{P_C}}}\frac{1}{\tau },
\end{equation}
yielding $\gamma  = 1$ for both canonical and grand canonical
ensembles.

Finally, we compute the exponent $\gamma$, Eq. (\ref{gamma}).
Putting $\tau=0$ in the expansion (\ref{EOS expansion}) leads to
\begin{equation}
P - {P_C} = \frac{{{P_C}h({v_c},{c_i},s)}}{{v_c^3}}{(v - {v_c})^3},
\end{equation}
which specifies $\delta=3$ for both the ensembles. So, the results
are the same as those computations from the mean field theory.

\section{Conclusion} \label{sec:conclusion}

We have investigated the thermodynamic properties of charged-AdS
TBHs in (dRGT) massive gravity coupled to PMI electrodynamic
source in the GCE. First, using the appropriate boundary
condition, we have fixed the electrostatic potential at the AdS
boundary and then written down all the thermodynamic quantities in
terms of the potential. We have evaluated the (finite) Euclidean
on-shell action and, by having it, we could obtain the
semi-classical partition function of TBHs. It was shown that the
obtained quantities extracted from the partition function satisfy
the first law of black hole thermodynamics in the extended phase
space. These quantities also obey the Smarr formula, Eq.
(\ref{Smarr}), in agreement with the method of scaling argument
developed by Kastor \cite{Kastor2009CQG,Kastor2010LovelockSmarr}.
As expected, our results, Eqs. (\ref{first law}) and
(\ref{Smarr}), proved that the variation of massive couplings
($c_i$) as well as cosmological constant ($\Lambda =-8 \pi P$) are
required for consistency of the extended first law of
thermodynamics with the Smarr formula.

Considering the grand canonical (fixed potential, $\Phi$)
ensemble, we have seen various thermodynamic phenomena, such as
vdW behavior in $d \ge 4$, RPT in $d \ge 5$, anomalous vdW
behavior in $d \ge 6$ and also triple points in $d \ge 6$
dimensions. As discussed in Sect. \ref{sec:comparison}, the RPT
phenomenon is observed only in the GCE and probably there does not
exist such a critical phenomenon in the canonical ensemble (but,
we emphasize that the existence of RPT phenomenon is not ruled out
completely in the canonical ensemble since our speculation has
relied on the numerical investigation). However, the rest of the
critical phenomena show up in the same dimensions in both
ensembles. In comparison to the TBHs in the GCE of Maxwell-massive
gravity \cite{Canonical-PMI-massive}, it has been observed that
the holographic critical phenomena in PMI-massive gravity start
appearing in one dimension lower. Interestingly, the critical
ratio up to two interaction potentials $O({\cal U}_2)$ has a
universal behavior as ${\rho _c} = \frac{{{P_C}{v_c}}}{{{{\tilde
T}_C}}} = \frac{{2s - 1}}{{4s}}$ in terms of the shifted Hawking
temperature. In $d \ge 6 $ dimensions, one can always set $s=2$ to
obtain the standard critical ratio of vdW fluid, $\rho_c =3/8$, in
higher dimensions.

The effects of massive graviton and PMI electromagnetic field
corrections are encoded in the deformation parameters $m_g$ and
$s$, respectively. The variation of the nonlinearity parameter
($s$) and also the graviton's mass ($m_g$) could produce or ruin a
specific phase transition. For example, in the limit $s \to 1$,
all the critical phenomena that presented in Sect. \ref{phase
transitions} disappear suddenly, but they can be found in higher
dimensions as we expected. In fact, in this case ($s \to 1$), the
outcomes of Maxwell-massive gravity are recovered
straightforwardly. In addition, for all the critical phenomena
except the vdW behavior in the present work, there is a lower
bound for the graviton's mass parameter ($m_g$), referred to as
$m_g^*$, in which no phase transition takes place for $m_g <
m_g^*$. In table \ref{tab:ensembles}, we have summarized the
results of critical phenomena in both ensembles for different
limits. We have also discussed black hole criticality in (Einstein
and massive) gravity coupled with conformally invariant Maxwell
source in Sect. \ref{sec:Massive-CIM}. In this limit, the final
results were presented in table \ref{tab:ensemblesCIM}. Finally,
we have computed the critical exponents in both ensembles and
found that they match the standard critical exponents of vdW
fluid. These are independent of spacetime dimensions as well.

There are several proposals to extend this work. One of the
possible extensions is to study other types of holographic phase
transitions which could appear only in $d \ge 7$ or $d \ge 8$
dimensions if one considers the higher order self-interaction
potentials, ${\cal U}_i$ (which means considering more massive
couplings ($c_i$) in the EoS). Furthermore, thermodynamic
stability properties of the obtained TBHs can be investigated by
introducing some thermodynamic coefficients such as isobaric
expansivity or adiabatic compressibility \cite{Dolan2014}. These
quantities can be realized in this paper from $T-r_+$ and $P-r_+$
diagrams, respectively. Also, following
\cite{Wei2015,Wei2019Mann}, it would be interesting to explore the
microscopic structure of massive-PMI TBHs from the thermodynamic
viewpoint by introducing the number density of the black hole
molecules. On the other hand, the exact massive-PMI TBHs
constructed here could potentially have applications in
gauge/gravity duality since massive couplings ($c_i$) in the
extended phase space are not fixed \textit{a priori}, so each
massive coupling has a specific interpretation on the gauge field
theory side. However, holographic interpretation of both the
massive couplings and thermodynamic phase transitions remains an
open question.

\begin{acknowledgments}
We are indebted to the referees for their constructive comments.
We wish to thank the Shiraz University Research Council. AD would
like to thank Soodeh Zarepour for helpful discussions.
\end{acknowledgments}


\begin{thebibliography} {0}


     \bibitem{CQG2017Review} D. Kubiznak, R.B. Mann and M. Teo, \textit{Black hole chemistry : thermodynamics with Lambda}, Class. Quant. Grav. \textbf{34} (2017) 063001.

    \bibitem{Karch2015}A. Karch and B. Robinson,\textit{ Holographic black hole chemistry}, JHEP \textbf{12} (2015) 01.

    \bibitem{KubiznakMann2012}D. Kubiznak and R.B. Mann, \textit{P-V criticality of charged AdS black holes}, JHEP \textbf{07} (2012) 033.

    \bibitem{Mann2012Gunasekaran}S. Gunasekaran, R.B. Mann and D. Kubiznak, \textit{Extended phase space thermodynamics for charged and rotating black holes and Born-Infeld vacuum polarization}, JHEP \textbf{11} (2012) 110.

    \bibitem{Mann2012Altamirano}N. Altamirano, D. Kubiznak, and R.B. Mann, \textit{Reentrant phase transitions in rotating anti-de Sitter black holes}, Phys. Rev. D \textbf{88} (2013) 101502.

    \bibitem{Mann2014TriplePoint}N. Altamirano, D. Kubiznak, R.B. Mann and Z. Sherkatghanad, \textit{Kerr-AdS analogue of triple point and solid/liquid/gas phase transition}, Class. Quant. Grav. \textbf{31} (2014) 042001.

    \bibitem{HennigarMann2017PRL}R.A. Hennigar and R.B. Mann, \textit{Superfluid Black Holes}, Phys. Rev. Lett. \textbf{118} (2017) 021301.

    \bibitem{Kastor2009CQG}D. Kastor, S. Ray, and J. Traschen, \textit{Enthalpy and the mechanics of AdS black holes}, Class. Quant. Grav. \textbf{26} (2009) 195011.

    \bibitem{Kastor2010LovelockSmarr}D. Kastor, D.S. Ray, and J. Traschen, \textit{Smarr formula and an extended first law for Lovelock gravity}, Class. Quant. Grav. \textbf{27} (2010) 235014.

    \bibitem{Kastor2018} D. Kastor, S. Ray, and J Traschen, \textit{Black hole enthalpy and scalar fields}, Class. Quant. Grav. \textbf{36} (2018) 024002.

    \bibitem{EmparanChamblin1999a}A. Chamblin, R. Emparan, C.V. Johnson, and R.C. Myers, \textit{Charged AdS black holes and catastrophic holography}, Phys. Rev. D \textbf{60} (1999) 064018.

    \bibitem{EmparanChamblin1999b}A. Chamblin, R. Emparan, C.V. Johnson, and R.C. Myers, \textit{Holography, thermodynamics and fluctuations of charged AdS black holes}, Phys. Rev. D \textbf{60} (1999) 104026.

    \bibitem{Caldarelli2000KerrNewmanAdS}M.M. Caldarelli, G. Cognola, and D. Klemm, \textit{Thermodynamics of Kerr-Newman-AdS black holes and conformal field theories}, Class. Quant. Grav. \textbf{17} (2000) 399.

    \bibitem{Fernando(2006)BI-AdS}S. Fernando, \textit{Thermodynamics of Born-Infeld-anti-de Sitter black holes in the grand canonical ensemble}, Phys. Rev. D \textbf{74} (2006) 104032.

    \bibitem{Mo2015NonExtendedLovelock} J-X. Mo and W-B. Liu, \textit{Non-extended phase space thermodynamics of Lovelock AdS black holes in the grand canonical ensemble}, Eur. Phys. J. C \textbf{75} (2015) 211.

    \bibitem{Cai2015}R.G. Cai, Y.P. Hu, Q.Y. Pan, and Y.L. Zhang, \textit{Thermodynamics of Black Holes in Massive Gravity}, Phys. Rev. D \textbf{91} (2015) 024032.

    \bibitem{Dehyadegari2020}A. Sheykhi, M. Arab, Z. Dayyani, and A. Dehyadegari, \textit{Alternative approach towards critical behavior and microscopic structure of the higher dimensional Power-Maxwell black holes}, Phys. Rev.D \textbf{101} (2020) 064019.

    \bibitem{Dolan2011CQG1}B.P. Dolan, \textit{The cosmological constant and the black hole equation of state}, Class. Quant. Grav. \textbf{28} (2011) 125020.

    \bibitem{Dolan2011CQG2}B.P. Dolan, \textit{Pressure and volume in the first law of black hole thermodynamics}, Class. Quant. Grav. \textbf{28} (2011) 235017.

    \bibitem{Dolan2011PRD}B.P. Dolan, \textit{Compressibility of rotating black holes}, Phys. Rev. D \textbf{84} (2011) 127503.


    \bibitem{Wei2012} S.W. Wei and Y.X. Liu, \textit{Triple points and phase diagrams in the extended phase space of charged Gauss-Bonnet black holes in AdS space}, Phys. Rev. D \textbf{90} (2014) 044057.

    \bibitem{Cai2013GaussBonnet}R.G. Cai, L.M. Cao, L. Li, and R.Q. Yang, \textit{P-V criticality in the extended phase space of Gauss-Bonnet black holes in AdS space}, JHEP \textbf{09} (2013) 005.

    \bibitem{HendiVahidinia2013}S.H. Hendi and M.H. Vahidinia, \textit{Extended phase space thermodynamics and P-V criticality of black holes with a nonlinear source}, Phys. Rev. D \textbf{88} (2013) 084045.

    \bibitem{Frassino2014}A.M. Frassino, D. Kubiznak, R.B. Mann, and F. Simovic, \textit{Multiple reentrant phase transitions and triple points in Lovelock thermodynamics}, JHEP \textbf{09} (2014) 080.

    \bibitem{MannGalaxies2014}N. Altamirano, D. Kubiznak, R.B. Mann, and Z. Sherkatghanad, \textit{Thermodynamics of rotating black holes and black rings: phase transitions and thermodynamic volume}, Galaxies \textbf{2} (2014) 89.

    \bibitem{DolanMann2014Lovelock}B.P. Dolan, A. Kostouki, D. Kubiznak, and R.B. Mann, \textit{Isolated critical point from Lovelock gravity}, Class. Quan. Grav. \textbf{31} (2014) 242001.

    \bibitem{BI-AdSBH-PV2014}D.-C. Zou, S.-J. Zhang, and B. Wang, \textit{Critical behavior of Born-Infeld AdS black holes in the extended phase space thermodynamics}, Phys. Rev. D \textbf{89} (2014) 044002.

    \bibitem{PV2014Lovelock} H. Xu, W. Xu, and L. Zhao, \textit{Extended phase space thermodynamics for third order Lovelock black holes in diverse dimensions}, Eur. Phys. J. C \textbf{74} (2014) 3074.

    \bibitem{PV2015LovelockBI-Belhaj}A. Belhaj, M. Chabab, H. El Moumni, K. Masmar, and M. B. Sedra, \textit{Ehrenfest scheme of higher dimensional AdS black holes in the third-order Lovelock-Born-Infeld gravity}, \textit{Int. J. Geom. Meth. Mod. Phys.} \textbf{12} (2015) 1550115.

    \bibitem{PVmassive2015PRD}J. Xu, L.M. Cao, and Y.P. Hu, \textit{P-V criticality in the extended phase space of black holes in massive gravity}, Phys. Rev. D \textbf{91} (2015) 124033.

    \bibitem{Hendi2017Mann-PRD}S.H. Hendi, , R.B. Mann, S. Panahiyan, and B. Eslam Panah, \textit{van der Waals like behavior of topological AdS black holes in massive gravity}, Phys. Rev. D \textbf{95} (2017) 021501(R).

    \bibitem{RPT-dRGTmassive-2017}D. Zou, R. Yue, and M. Zhang, \textit{Reentrant phase transitions of higher-dimensional AdS black holes in dRGT massive gravity}, Eur. Phys. J. C \textbf{77} (2017) 256.

    \bibitem{PV-BlackBranes-2017Hennigar}R.A. Hennigar, \textit{Criticality for charged black branes}, \textit{JHEP} \textbf{09} (2017) 082.

    \bibitem{Mir2019a} M. Mir, R.A. Hennigar, J. Ahmed and R.B. Mann, \textit{Black hole chemistry and holography in generalized quasi-topological gravity}, JHEP \textbf{08} (2019) 068.

    \bibitem{Mir2019b}M. Mir and R.B. Mann, \textit{On generalized quasi-topological cubic-quartic gravity: thermodynamics and holography}, JHEP \textbf{07} (2019) 012.

    \bibitem{EPJC2019} S.H. Hendi, A. Dehghani, \textit{Criticality and extended phase space thermodynamics of AdS black holes in higher curvature massive gravity}, Eur. Phys. J. C \textbf{79} (2019) 227.

   \bibitem{PRD2020}A. Dehghani, S.H. Hendi, and R.B. Mann, \textit{Range of novel black hole phase transitions via massive gravity: Triple points and N-fold reentrant phase transitions}, Phys. Rev. D \textbf{101} (2020) 084026.

    \bibitem{CQG2020}A. Dehghani, S.H. Hendi, \textit{Charged black hole chemistry with massive gravitons}, Class. Quant. Grav. \textbf{37} (2020) 024001.

    \bibitem{PV-2020-Triple-MassiveGravity}B. Liu, Bo, Z-Y. Yang, and R-H. Yue, \textit{Tricritical point and solid/liquid/gas phase transition of higher dimensional AdS black hole in massive gravity}, Ann. Phys. \textbf{412} (2020) 168023.

    \bibitem{PV-2019-Einstein-Horndeski}Y-P. Hu, H-A. Zeng, Z-M Jiang, and H. Zhang,\textit{ P-V criticality in the extended phase space of black holes in Einstein-Horndeski gravity}, Phys. Rev. D \textbf{100} (2019) 084004.

    \bibitem{PV-2020-BraneworldBHs}M.U. Shahzad and M.Z. Ashraf, \textit{P-V criticality in braneworld black holes with logarithmic-corrected entropy}. Mod. Phys. Lett. A \textbf{35} (2020) 2050099.

    \bibitem{Gue2020}X-Y. Guo, H-F. Li, L-C. Zhang, and R. Zhao, \textit{Continuous phase transition and microstructure of charged AdS black hole with quintessence}, Eur. Phys. J. C \textbf{80} (2020) 1.

    \bibitem{Hinterbichler2012Review}K. Hinterbichler, \textit{Theoretical Aspects of Massive Gravity}, Rev. Mod. Phys. \textbf{84} (2012) 671.

    \bibitem{deRhamREVIEW2014}C. de Rham, \textit{Massive gravity}, Living Rev. Rel. \textbf{17} (2014) 7.

    \bibitem{Schmidt-May2016a-DM} E. Babichev, L. Marzola, M. Raidal, A. Schmidt-May, F. Urban, H. Veermäe, and M. von Strauss. \textit{Bigravitational origin of dark matter}, Phys. Rev. D \textbf{94} (2016) 084055.

    \bibitem{Schmidt-May2016b-DM} E. Babichev, L. Marzola, M. Raidal, A. Schmidt-May, F. Urban, H. Veermäe, and M. von Strauss, \textit{Heavy spin-2 dark matter}, J. Cosmol. Astropart. Phys.\textbf{09} (2016) 016.

    \bibitem{Akrami2013}Y. Akrami, T.S. Koivisto and M. Sandstad, \textit{Accelerated expansion from ghost-free bigravity: a statistical analysis with improved generality}, JHEP \textbf{03} (2013) 99.

    \bibitem{Akrami2015} Y. Akrami, S.F. Hassan, F. Könnig, A. Schmidt-May, and A.R. Solomon, \textit{Bimetric gravity is cosmologically viable}, Phys. Lett. B \textbf{748} (2015) 37.

    \bibitem{deRham2010Gabadadze}C. de Rham and G. Gabadadze, \textit{Generalization of the Fierz-Pauli Action}, Phys. Rev. D \textbf{82} (2010) 044020.

    \bibitem{dRGT2011} C. de Rham, G. Gabadadze, and A.J. Tolley, \textit{Resummation of Massive Gravity}, Phys. Rev. Lett. \textbf{106} (2011) 231101.

    \bibitem{BDghost1972}D.G. Boulware and S. Deser, \textit{Can gravitation have a finite range?}, Phys. Rev. D \textbf{6} (1972) 3368.

    \bibitem{HassanRosen2012PRL}S.F. Hassan and R.A. Rosen, \textit{Resolving the ghost problem in non-linear massive gravity}, Phys. Rev. Lett. \textbf{108} (2012) 041101.

    \bibitem{vDVZ1970a}H. van Dam and M. Veltman, \textit{Massive and mass-less Yang–Mills and gravitational fields}. Nucl. Phys. B \textbf{22} (1970) 397.

    \bibitem{vDVZ1970b}V. Zakharov, \textit{Linearized gravitation theory and the graviton mass}. JETP Lett. \textbf{12} (1970) 312.

    \bibitem{FierzPauli1939}M. Fierz and W. Pauli, \textit{On relativistic wave equations for particles of arbitrary spin in an electromagnetic field}, Proc. Roy. Soc. Lond. A \textbf{173} (1939) 211.

    \bibitem{MG2012JHEP} S.F. Hassan, R.A. Rosen, and A. Schmidt-May, \textit{Ghost-free massive gravity with a general reference metric}, JHEP \textbf{02} (2012) 026.

   \bibitem{Alberte2013PRD} L. Alberte and A. Khmelnitsky, \textit{Reduced massive gravity with two Stückelberg fields}, Phys. Rev. D \textbf{88} (2013) 064053.

    \bibitem{Alberte2015PRD} L. Alberte and A. Khmelnitsky, \textit{Stability of massive gravity solutions for holographic conductivity}, Phys. Rev. D \textbf{91} (2015) 046006.

    \bibitem{ReducedMG-2016-GhostFree}H. Zhang and X-Z. Li, \textit{Ghost free massive gravity with singular reference metrics}, Phys. Rev. D \textbf{93} (2016) 124039.

    \bibitem{TQDo2016massive}T.Q. Do, \textit{Higher dimensional nonlinear massive gravity}, Phys. Rev. D \textbf{93} (2016) 104003.

    \bibitem{TQDo2016bigravity}T.Q. Do, \textit{Higher dimensional massive bigravity}, Phys. Rev. D \textbf{94} (2016) 044022.

    \bibitem{MassiveCosmologies2011}G. D'Amico, C. de Rham, S. Dubovsky, G. Gabadadze, D. Pirtskhalava, and A. J. Tolley, \textit{Massive cosmologies}, Phys. Rev. D \textbf{84} (2011) 124046

    \bibitem{Chamseddine2011Volkov}A.H. Chamseddine and M.S. Volkov, \textit{Cosmological solutions with massive gravitons}, Phys. Lett. B \textbf{704} (2011) 652.

    \bibitem{deRham2017}C. de Rham, J.T. Deskins, A.J. Tolley, and S.Y. Zhou, \textit{Graviton mass bounds}, Rev. Mod. Phys. \textbf{89} (2017) 025004.

    \bibitem{Koyama2011(PRL)}K. Koyama, G. Niz, and G. Tasinato, \textit{Analytic solutions in nonlinear massive gravity}, Phys. Rev. Lett. \textbf{107} (2011) 131101.

    \bibitem{Nieuwenhuizen2011(PRD)}T.M. Nieuwenhuizen, \textit{Exact Schwarzschild-de Sitter black holes in a family of massive gravity models}, Phys. Rev. D \textbf{84} (2011) 024038.

    \bibitem{Gruzinov2011Mirbabayi}A. Gruzinov and M. Mirbabayi, \textit{Stars and Black Holes in Massive Gravity}, Phys. Rev. D \textbf{84} (2011) 124019.

    \bibitem{Berezhiani2012} L. Berezhiani, G. Chkareuli, C. de Rham, G. Gabadadze, and A. J. Tolley, \textit{On black holes in massive gravity}, Phys. Rev. D \textbf{85} (2012) 044024.

    \bibitem{Rosen2017BHs}R.A. Rosen, \textit{Non-singular black holes in massive gravity: time-dependent solutions}, JHEP \textbf{10} (2017) 206.

    \bibitem{Vegh2013} D. Vegh, \textit{Holography without translational symmetry}, arXiv:1301.0537.

    \bibitem{BlakeTong2013} M. Blake and D. Tong, \textit{Universal resistivity from holographic massive gravity}, Phys. Rev. D \textbf{88} (2013) 106004.

    \bibitem{Davison2013} R.A. Davison, \textit{Momentum relaxation in holographic massive gravity}, Phys. Rev. D \textbf{88} (2013) 086003.

    \bibitem{Baggioli2015PRL}M. Baggioli and Oriol Pujolas, \textit{Electron-phonon interactions, metal-insulator transitions, and holographic massive gravity}, Phys. Rev. Lett. 114 (2015) 251602.

    \bibitem{Baggioli2016JHEP} L. Alberte, M. Baggioli, A. Khmelnitsky, and O. Pujolas, \textit{Solid holography and massive gravity}, JHEP \textbf{02} (2016) 114.

    \bibitem{Baggioli2018PRL} L. Alberte, M. Ammon, A. Jim´enez-Alba, M. Baggioli, and
    Oriol Pujolas, \textit{Holographic phonons}, Phys. Rev. Lett. \textbf{120} (2018) 171602.

    \bibitem{HassaineMartinez2007} M. Hassaine and C Martinez, \textit{Higher-dimensional black holes with a conformally invariant Maxwell source}, Phys. Rev. D \textbf{75} (2007) 027502.

    \bibitem{HassaineMartinez2008} M. Hassaine and C. Martinez, \textit{Higher-dimensional charged black hole solutions with a nonlinear electrodynamics source}, Class. Quant. Grav. \textbf{25} (2008) 195023.

    \bibitem{Canonical-PMI-massive}A. Dehghani, S.H. Hendi, \textit{Topological black hole chemistry in massive gravity with power-Maxwell invariant field}, submitted for publication

   \bibitem{Dehyadegari2017MGPMI}A. Dehyadegari, M. Kord Zangeneh, and A. Sheykhi, \textit{Holographic conductivity in the massive gravity with power-law Maxwell field}, Phys. Lett. B \textbf{773} (2017) 344.

    \bibitem{Hendi2009Sedehi}S.H. Hendi and H.R. Rastegar-Sedehi, \textit{Ricci flat rotating black branes with a conformally invariant Maxwell source}, Gen. Rel. Grav. \textbf{41} (2009) 1355.

    \bibitem{Hendi2010PMI} S.H. Hendi, \textit{The relation between F (R) gravity and Einstein-conformally invariant Maxwell source}, Phys. Lett. B \textbf{690} (2010) 220.

    \bibitem{Heisenberg1936} W. Heisenberg and H. Euler, \textit{Consequences of Dirac's theory of positrons}, Z. Phys. \textbf{98} (1936) 714.

    \bibitem{NED-QED-a}V.A. De Lorenci, R. Klippert, M. Novello, and J.M. Salim, \textit{Light propagation in non-linear electrodynamics}, Phys. Lett. B \textbf{482} (2000) 134.

    \bibitem{NED-QED-b}P. Gaete, Patricio, and J. Helayel-Neto, \textit{Remarks on nonlinear electrodynamics}, Eur. Phys. J. C \textbf{74} (2014) 3182.

    \bibitem{NED-QED-c}T.C. Adorno, D.M. Gitman, and A.E. Shabad, \textit{Coulomb field in a constant electromagnetic background}, Phys. Rev. D \textbf{93} (2016) 125031.

    \bibitem{Peskin}M.E. Peskin and D.V. Schroeder, \textit{An introduction to quantum field theory}, Addison-Wesley, Reading, U.S.A. (1995).

    \bibitem{Hendi2012} S.H. Hendi, \textit{Asymptotic charged BTZ black hole solutions}, JHEP \textbf{03} (2012) 65.

    \bibitem{EFT-NED-QED}G.V. Dunne, \textit{Heisenberg-Euler effective Lagrangians: basics and extensions}, From Fields to Strings: Circumnavigating Theoretical Physics: Ian Kogan Memorial Collection (In 3 Volumes) (2005) 445.

    \bibitem{Shabad2015}C.V. Costa,  D.M. Gitman, and A.E. Shabad, \textit{Finite field-energy of a point charge in QED}, Physica Scripta \textbf{90} (2015) 074012.

    \bibitem{Jackson}J.D. Jackson, \textit{Classical Electrodynamics}, third edition, Wiley, New York U.S.A. (1998).

    \bibitem{Fouch}M. Fouch, R. Battesti, and C. Rizzo, \textit{Limits on nonlinear electrodynamics}, Phys. Rev. D \textbf{93} (2016) 093020.


    \bibitem{Eslampanah2021} B. Eslam Panah, \textit{Can the power Maxwell nonlinear electrodynamics theory remove the singularity of electric field of point-like charges at their locations?}, arXiv: 2103.08343 (accepted in EPL).

    \bibitem{Gross1987}D.J. Gross and J.H. Sloan, \textit{The quartic effective action for the heterotic string}, Nuc. Phys. B \textbf{291} (1987) 41.

    \bibitem{ZwiebachStringBook}B. Zwiebach, \textit{A first course in string theory}, Cambridge university press, 2004.


    \bibitem{AdS/CFT-PMI-a}J. Jing, Q. Pan, and S. Chen, \textit{Holographic superconductors with Power-Maxwell field}, JHEP \textbf{11} (2011) 45.

    \bibitem{AdS/CFT-PMI-b}D. Roychowdhury, \textit{AdS/CFT superconductors with Power Maxwell electrodynamics: reminiscent of the Meissner effect}, Phys. Lett. B \textbf{718} (2013) 1089.

    \bibitem{AdS/CFT-PMI-c}J. Jing, L. Jiang, and Q. Pan, \textit{Holographic superconductors for the Power?Maxwell field with backreactions}, Class. Quant. Grav. \textbf{33} (2015) 025001.


    \bibitem{AdS/CFT-PMI-d} B. Mu, Benrong, P. Wang, and H. Yang, \textit{Holographic DC conductivity for a power-law Maxwell field}, Eur. Phys. J. C \textbf{78} (2018) 1005.

    \bibitem{Shabad2011} A.E. Shabad and V.V. Usov, \textit{Effective Lagrangian in nonlinear electrodynamics and its properties of causality and unitarity}, Phys. Rev. D \textbf{83} (2011) 105006.

    \bibitem{GibbonsHawking1977}G.W. Gibbons and S.W. Hawking, \textit{Action integrals and partition functions in quantum gravity}, Phys. Rev. D \textbf{15} (1977) 2752.

    \bibitem{HawkingPage1983}S. Hawking and D.N. Page, \textit{Thermodynamics of black holes in Anti-de Sitter space}, Commun. Math. Phys. \textbf{87} (1983) 577.

    \bibitem{Hawking1975}S.W. Hawking, \textit{Particle creation by black holes}, Commun. Math. Phys. \textbf{43} (1975) 199.

    \bibitem{BardeenCarterHawking1973}J.M. Bardeen, B. Carter, and S.W. Hawking, \textit{The four laws of black hole mechanics}, Commun. Math. Phys. \textbf{31} (1973) 161.

    \bibitem{Ashtekar1984AMD}A. Ashtekar and A. Magnon, \textit{Asymptotically anti-de Sitter space-times}, Class. Quant. Grav. \textbf{1} (1984) L39.

    \bibitem{Ashtekar2000AMDmass}A. Ashtekar and S. Das, \textit{Asymptotically anti-de Sitter spacetimes: conserved quantities}, Class. Quant. Grav. \textbf{17} (2000) L17.

   \bibitem{KlauberBook} R.D. Klauber, \textit{Student Friendly Quantum Field Theory: Basic Principles and Quantum
    Electrodynamics} (Fairfield, IA: Sandtrove Press)

    \bibitem{Zee2010-book}A. Zee, \textit{Quantum field theory in a nutshell}, Princeton, NJ: Princeton university press (2010).

    \bibitem{Natsuume2015-book}M. Natsuume, \textit{AdS/CFT duality user guide}, Berlin: Springer (2015)


    \bibitem{York1990GrandCan}H.W. Braden, J.D. Brown, B.F. Whiting, and J.W. York Jr., \textit{Charged black hole in a grand canonical ensemble}, Phys. Rev. D \textbf{42} (1990) 3376.

    \bibitem{Witten1998b}E. Witten, \textit{Anti-de Sitter space, thermal phase transition and confinement in gauge theories}, Adv. Theor. Math. Phys. \textbf{2} (1998) 505.

    \bibitem{Mann2017Sinamuli} M. Sinamuli and R.B. Mann, \textit{Higher order corrections to holographic black hole chemistry}, Phys. Rev. D \textbf{96} (2017) 086008.

    \bibitem{vanderWaals1873}J.D. van der Waals, \textit{Over de continuiteit van den gas-en vloeistoftoestand (On the continuity of the gaseous and the liquid state)}, Dessertation, Leiden (1873).

    \bibitem{Hudson1904} C.S. Hudson, \textit{Die gegenseitige loslichkeit von nikotin in wasser}, Zeit. Phys. Chem. \textbf{47} (1904) 113.

    \bibitem{Narayanan1994}T. Narayanan and A. Kumar, \textit{Reentrant phase transitions in multicomponent liquid mixtures}, Phys. Rep. \textbf{249} (1994) 135.

    \bibitem{LiquidCrystalsBook} S. Singh and D.A. Dunmur, Liquid Crystals: Fundamental 52 (World Scientific, Singapore, 2002).

    \bibitem{Huang2009-book}K. Huang, \textit{Introduction to statistical physics}, \textit{CRC press} (2009).

    \bibitem{Kaputsa2006-book}J.I. Kapusta and C. Gale, \textit{Finite-temperature field theory: Principles and applications}, \textit{Cambridge University Press} (2006).

    \bibitem{Zinn-Justin1996-book}J. Zinn-Justin, \textit{Quantum field theory and critical phenomena}, \textit{Clarendon Press} (1996).

    \bibitem{Dolan2014}B.P. Dolan, \textit{Thermodynamic stability of asymptotically anti-de Sitter rotating black holes in higher dimensions}, Class. Quant. Grav. \textbf{31} (2014) 165011.

    \bibitem{Wei2015}S.W. Wei and Y.-X. Liu, \textit{Insight into the microscopic structure of an AdS black hole from a thermodynamical phase transition}, Phys. Rev. Lett. \textbf{115} (2015) 111302.

    \bibitem{Wei2019Mann}S.W. Wei, Y.-X. Liu, and R.B. Mann, \textit{Repulsive Interactions and Universal Properties of Charged Anti-de Sitter Black Hole Microstructures}, Phys. Rev. Lett. \textbf{123}, (2019) 071103.




\end{thebibliography}
\end{document}